\documentstyle[aps,preprint,epsf]{revtex}
\tightenlines
\newcounter{abc}
\renewcommand{\theequation}{\arabic{equation}\alph{abc}}
\begin{document}

\title{Lattice-switch Monte Carlo}
\author{A.D. Bruce, A.N. Jackson, G.J. Ackland \& N.B. Wilding}
\address{Department of Physics and Astronomy, The University of Edinburgh\\
Edinburgh, EH9 3JZ, Scotland, United Kingdom}
\maketitle

\begin{abstract}

We present a Monte Carlo method for the direct evaluation of the difference
between the free energies of two crystal structures. The method is built on a
lattice-switch transformation that maps a configuration of one structure onto a
candidate configuration of the other by `switching' one set of lattice vectors
for the other, while keeping the displacements with respect to the lattice sites
constant. The sampling of the displacement  configurations is biased,
multicanonically, to favor paths leading to {\em gateway} arrangements for which
the Monte Carlo switch to the candidate configuration will be accepted. The
configurations of both structures can then be efficiently sampled in a single
process, and the difference between their free energies evaluated from their
measured probabilities.  We explore and exploit the method in the context of
extensive studies of systems of hard spheres. We show that the efficiency of the
method is controlled by the extent to which the switch conserves correlated
microstructure.  We also show how, microscopically, the procedure works: the
system finds  gateway  arrangements which fulfill the sampling bias
intelligently.   We establish, with high precision, the differences between the
free energies of the two close packed structures ({\it fcc} and {\it hcp}) in
both the constant density and the  constant pressure ensembles.

\end{abstract}

\section{Introduction \label{sec:intro}}

Let us pose the problem. We are presented with a material whose chemical
composition is known; we are provided  with a model of the interatomic
interactions; and we have identified  two candidate crystalline structures. 
How should we proceed to determine which structure will be favored under given
conditions?  Of course, equilibrium statistical mechanics tells us what we must
do, in principle: the favored structure will be that which has the {\it greater 
a priori probability} or {\it configurational weight}; or, in equivalent
thermodynamic parlance, {\it lower free energy}.  Thus the task is to {\em
compare} the configurational weights of (determine the {\em difference} between
the free energies of)  the candidate structures. 

A variety of approximate strategies exist for addressing this problem
\cite{jacucciquirke}.  But it is clear that if one desires a  technique that is
both generally applicable and reliable (that is, has quantifiable uncertainties)
one must look to the Monte Carlo (MC)  method \cite{ormoldyn}, the standard
computational tool for dealing with many-body systems \cite{binrev}.

The application of MC methods to the study of phase-behavior  presents a 
generic problem \cite{frenkelrev,mikeallanrev}: the free energy of a phase cannot
be expressed (in a practically useful form) as a canonical average over the
associated configurations;  free-energy-estimation inevitably entails
simulations that visit a substantially wider spectrum of configurations, which
together form a {\em path} through configuration space \cite{howtolabelpath}. The strategic choices to
be made concern the path itself ---ultimately, the physical character of
the additional configurations sampled--- 
and the sampling procedure. 

An acceptable path will fall into one or other of two categories --- we will
call them {\em reference-state} and {\em inter-phase} paths. A reference-state
path links (comprises sets of configurations that interpolate between)  the
configuration space associated with each phase to  the 
configuration space
associated with some reference system \cite{ortwo} whose free energy is known.
An inter-phase path links the  configuration space of one phase to
that of the other.  Both categories of path embrace many further 
sub-categories.  Thus, a reference-state-path may run through a  space of
thermodynamic coordinates  or through a  space of model parameters. An
inter-phase path may  be  `physically-motivated' ---modeling authentically
the configurations through which a system actually passes
in the course of a phase transformation;
or  it may be `computationally-motivated' (`non-physical')
---designed, pragmatically, to deliver a result.

The sampling procedures used to explore the chosen path also fall, broadly, into
one or other of two categories -- we will call them {\em multi-stage} and {\em
single-stage}. The multi-stage approach entails a {\em series} of  independent
simulations each of which explores a different  point  on the path; the
simulations may determine simply the derivative of the free energy at each point
(the {\em integration method}, IM),  or the difference between the free energies
of adjacent points  (the {\em overlap method}).  The single-stage approach
involves, in essence {\em one} simulation exploring the {\em entire} path.

There are very many ways in which one can respond to these strategic choices.
Many of them are represented in the large literature devoted to this
problem \cite{examples}-\cite{mauhuse}. 
But all of them, in our view, lack one or
more of the characteristics (generality, transparency, precision) of a
definitively-satisfactory solution to such a fundamental and simply-posed
problem. 

In seeking that solution it seems to us there are good {\it a priori} grounds
for  favoring an {\em inter-phase} path, explored by {\em single-stage
sampling}.   The prejudice on the choice of path reflects the fact that,
in using
a reference state path, one has to determine, separately, the  absolute  free
energies of each phase. These absolute
free energies are typically very large --{\em arbitrarily} so
in the vicinity of a phase boundary--  compared to the quantity (their
difference) which is actually of interest. In contrast,
using an inter-phase path allows one
to focus directly on this quantity. The {\it a priori} preference for a
single-stage sampling rests on the transparency with which the associated
uncertainties (error bounds) are prescribed. We shall return to these points in
section~\ref{SEC:PROSPECTS}. With these strategic choices made, one is left
with two tasks ---one conceptual (designing the inter-phase path), and the other
practical (formulating the sampling algorithm).

The practical issue is relatively easily addressed. In recent years,  the Monte
Carlo toolkit has been significantly enhanced to provide a range of extended
sampling techniques --- multicanonical \cite{bn}, expanded-ensemble \cite{lyu} and simulated-tempering
\cite{marinariparisi}.  
These methods (whose
origins can be traced back to much earlier pioneering work \cite{pioneering})
allow one to construct a MC procedure that will traverse virtually any desired
path through configuration space. Here we adopt the multicanonical framework. In this framework, the desired path is represented as a
discrete series of macrostates, defined by some  chosen macroscopic property
\cite{howtolabelpath}; in the multicanonical sampling procedure each macrostate
is visited with a probability that is enhanced, or diminished, with respect to
its canonical value, by an amount that is controlled by  a {\em multicanonical
weight}; the set of weights is constructed so that, while the canonical
probabilities vary vastly over the path, the multicanonical probabilities are
essentially constant, allowing the whole path to be negotiated in one
simulation. 

The core issue is, then, the {\em design} of the inter-phase-path ---at heart,
the choice of an appropriate {\em order parameter}
\cite{callitorderparameteter}. The choice is important: it determines, implicitly 
\cite{notobviousapriori}, the  nature of the configurations sampled during the  
inter-phase traverse, the MC-time required for that traverse, and thence the
statistical quality of the final results.  

Outside the context of {\em structural}-phase behavior --in the case of
liquid-gas phase behavior for example ---the choice is clear and a
multicanonical strategy is securely in place. The order-parameter is identified
with that ---the density--- associated with the accompanying critical point. 
The resulting inter-phase configurations are then generically {\em
inhomogeneous}, comprising two coexisting regions (one of each phase), separated
by an interface. On this path, it is the free energy cost of this interface that
provides the ergodic barrier which has to be surmounted  by multicanonical
weighting  \cite{interface}.  The passage along the path (the motion of the
interface) involves processes which differ only in scale from those already
represented in the microscopic dynamics of a single phase. This approach is
illustrated schematically in  Fig.~(\ref{fig:schematic}a). It has been
successfully used in studies of phase behavior in ferromagnets \cite{bergint},
fluids \cite{nbw} and lattice gauge theories \cite{lgt}.

In the context of {\em structural} phase behavior it is clear that this kind of 
strategy will not usually be fruitful \cite{exceptions}. In such systems a
traverse through an inhomogeneous two-phase (necessarily non-crystalline) region 
will involve substantial, physically slow, restructuring --- vulnerable to
further ergodic traps, and compounding the intrinsic slowness of the
multicanonical sampling process. To the two general {\em a priori}  preferences
expressed above we thus add a third, specific to the structural context: {\em
the inter-phase path should comprise macrostates that are single-phase, and 
crystalline}. This paper shows how to identify, build and exploit a path of
this kind.
 
The key  ideas are simple. In any crystalline configuration each atomic position
coordinate may be expressed as the sum of a {\em lattice vector} and a {\em
displacement vector}. The configuration space associated with each structure,
individually, may be explored by standard MC procedures which stochastically 
update the displacement vectors while keeping the lattice vectors constant.  In
{\em principle} the passage from one phase to the other may be accomplished by a
{\em lattice switch} (LS) in which one  entire set  of lattice vectors is
replaced with the other, while the displacement vectors  are held fixed. 
Formally this LS can be incorporated  into the MC procedure simply by treating
the lattice type as an additional  stochastic variable.  In practice {\em this}
inter-phase `path' (blind-leap) will not work. Implemented this way the LS will
map a `typical' configuration of one structure 
onto an `untypical' (high-energy)
conjugate configuration \cite{conjugatedef}; 
the associated MC step will generally  be rejected.
To make it work the LS needs to be extended to include two segments of `path'
(each lying entirely within one phase) which connect  the sets of equilibrium
configurations with the special configurations 
(we will call them {\em gateway configurations})  
from which a successful LS can be initiated
\cite{gatewaystatedef}. These path-segments may be labeled by an
`order-parameter'  which   measures the mismatch between the energies
of the configurations linked by LS.
This order parameter has a high value for
the equilibrium configurations, lying at one end of a path segment:
these configurations are not energy-matched \cite{energymatcheddef} 
to their conjugates.
It has  a low value for the gateway configurations at the other end: 
gateway configurations
(whatever other attributes they may have) are necessarily energy-matched
to their conjugates. 
Multicanonical weights are attached to
the macrostates of this order parameter, so that the multicanonical sampling
procedure explores both path segments evenly, surmounting the probabilistic
barrier which, in this case, reflects the {\em smallness} of the statistical
weight of the gateway configurations. Together, the multicanonical sampling and
the lattice switch provide a configuration-space `look and leap'
(Fig.~\ref{fig:schematic}b) which  visits both  phases while remaining at all
times crystalline.

The LS method was introduced by us and described in outline form in an earlier
brief communication \cite{lsprl}. Since that time it has been applied by two
other groups \cite{mauhuse,pronkfrenkel}. The present paper has three principal
objectives. 

The first objective (with which we have already engaged in the preceding
discussion) is to explain the core idea more fully: the `idea' (biased sampling
to facilitate a global coordinate change) represents, we believe, a significantly
new form of extended sampling, which merits further exposure.

Our second objective is to achieve a deeper understanding of  how the process
works --in particular the implications of the
{\em form} chosen for the LS operation
adopted (it is not unique) and the microscopic character of the  gateway
configurations which the system locates in response to the multicanonical weighting,
tailored to support that operation. We show that the efficiency of the LS 
operation depends significantly on the extent to which it conserves correlated
microstructure. And we find that the 
gateway configurations have features which reflect
the specific nature of the lattice-switch transformation we adopt,
in a microscopically intelligible (even intelligent) way.

Our third objective is to extend our study of the  phase-behavior of
hard-spheres.  This problem  is of enduring interest, displaying a richness that
belies the simplicity of the model itself \cite{pnphsrev}. The relative
stability of the two closed-packed ({\it fcc} and {\it hcp}) structures is
particularly finely balanced: the entropy difference \cite{entropyhere} is so
small (smaller than the entropy change at freezing by of order  $10^{-3}$) that
it can easily be lost in statistical  uncertainties. Discrepancies (large, in
relative terms)  between a recent  IM study \cite{woodcock} and its predecessors
\cite{frenkelladd}  provided the motivation for our development of the LS
method. In this paper we present results in  the constant-density ensemble, both
near the melting density and  at the close-packed limit. In so doing we resolve
the discrepancy  between the results, near melting, reported in our initial
study \cite{lsprl}  and those  ---also using LS--- reported recently by Pronk \&
Frenkel \cite{pronkfrenkel}: the fault was ours, stemming from a failure to
recognize the consequences of
center-of-mass drift. We also show that the method can be extended
straightforwardly --in this case at least-- to the constant-pressure ensemble. 

The paper is structured as follows. Section~\ref{SEC:FORMULATION}  sets out the theoretical
framework. We define the  model, the competing structures,  and the associated
configurational weights: in the case of hard sphere systems the latter are
purely entropic. We identify an appropriate form of lattice-switch
transformation: here, it is designed to capitalize on the  close-packed layers
common to both structures. To bias the displacement sampling we need to define
an appropriate measure of the `energy cost' of the lattice switch; we will see
that the number of pairs of overlapping spheres created by the transformation
fulfills this role simply and effectively. The efficiency of the method also
potentially
depends on the choice of representation of both the lattice-to-lattice mapping
and the particle displacements: we discuss the principles involved in the choice
of representation.
Section ~\ref{SEC:IMPLEMENTATION} provides computational
implementation details, including the procedures used to evolve
an appropriate multicanonical sampling distribution. Section ~\ref{SEC:RESULTS} contains 
our results. Finally, in section~\ref{SEC:PROSPECTS}, we offer our conclusions in relation to both
the hard sphere system and the lattice-switch method.

\section{Formulation \label{sec:formulation}}

\subsection{The model system \label{subsec:model}}

We consider a system of $N$ particles, of spatial coordinates $\left\{\vec{r}\right\}$, confined
within  a volume $V$, and subject to periodic boundary conditions. The
interactions are those of hard spheres of diameter $D$; the configurational
energy is of the form 
\begin{equation} 
\label{eq:hsinteraction} 
E(\left\{\vec{r}\right\}) = \left\{\begin{array}{ll} 
0 & \mbox{if } r_{ij} \ge D \,\forall i,j \\ 
\infty & \mbox{otherwise}
\end{array}
\right.
\end{equation}
where $r_{ij}= \mid \vec{r}_i -\vec{r}_j\mid$. 
The total configurational weight associated with this system is
\begin{equation} 
\label{eq:totconfwt}
\Omega (N, V) = \prod_i [ \int_V  d\vec r _i ]\prod_{<ij>} \Theta (r_{ij} -D )
\end{equation}
where $\Theta (x) \equiv 1\,\, (0)$ for $ x \ge 0\,\, (< 0)$, and the product on
$<ij>$ extends over all particle pairs. The associated entropy
density is
\begin{equation}
\label{eq:totent}
s(N,V) \equiv \frac{1}{N} \ln \Omega (N, V ) 
\end{equation}

We are concerned with the entropy  of specific phases (the two familiar
crystalline close-packed structures) of this system. In general, the entropy of a phase
measures the weight of the configurations satisfying some constraint that is
characteristic of that phase. It is therefore necessary in principle (although
in practice the issue is typically skirted) to formulate a
constraint that identifies a configuration as `belonging to' a given crystalline
phase. One can do so --very naturally, and in the traditions of lattice dynamics
\cite{bornhuang}--  by decomposing the particle position coordinates into a
sum of `lattice' and `displacement' vectors:
\begin{equation}
\label{eq:decomposition}
\vec{r}_i=\vec{R}^{\alpha}_i+\vec{u}_i
\end{equation} 
Here $\{\vec{R}\}_\alpha \equiv\vec{R}^{\alpha}_i, i=1\ldots N$ 
is a set of fixed (configuration-independent) vectors associated 
with the crystalline structure labeled $\alpha$. We will refer to them as 
`lattice vectors'. But we use this term a little loosely: more precisely, we 
mean the set of vectors identified by the orthodox crystallographic lattice, 
convolved with the orthodox basis \cite{notbravaislattice}. The other
vectors, $\left\{\vec{u}\right\}$ , represent displacements with respect to the `lattice' sites; 
the symmetry of the structures of interest here ensures that these displacements
have zero ensemble average. This framework provides us with a number of ways of
identifying the configurations to be associated with structure $\alpha$.
First, one might adopt the criterion that
all particle displacements, with respect to the associated lattice sites, 
lie within some nominated {\em spatial} cutoff, chosen to be sufficiently large
that the results are independent of its specific value.
This criterion has the merit that it does not stray beyond the concepts of equilibrium statistical mechanics.
Alternatively one might identify the relevant configurations as the set that
are accessible from {\em some} member of the set (the perfect crystalline
state, for example) within some nominated {\em temporal} cutoff.
The merit of {\em this} choice is that it is a quasi-formal expression
of what, in practice, computer simulation attacks on this problem actually {\em do},
albeit implicitly: the free energy assigned to a phase (in, for example,
IM-based studies) represents the weight of the configuration space sampled
on the time scale of the simulation. The result should be independent of that time
scale provided it (the scale) is long enough that the configuration space of each structure
is effectively sampled, but still short compared to inter-phase
crossing times. Whichever view one takes (in practice we adopt the latter: 
see section~\ref{subsec:MCprocedures}) one
may write, for the configurational weight associated with structure $\alpha$
\begin{equation} 
\label{eq:partconfwt}
\Omega (N, V, \alpha) = \prod_i [ \int_{\framebox{$\alpha$}}  d\vec u _i ] 
\prod_{<ij>}\Theta (r_{ij} -D )
\end{equation}
where $\int_{\framebox{$\alpha$}}$ signifies integration subject to the chosen
configurational constraint.

In the thermodynamic ($N\rightarrow \infty$) limit, the associated entropy
density
\begin{equation}
\label{eq:partent}
s(N,V,\alpha) \equiv \frac{1}{N} \ln \Omega (N, V, \alpha ) 
\end{equation}
depends only on the particle number density, which we write in the
dimensionless form 
\begin{equation}
\label{eq:densitydef}
\tilde{\rho} \equiv \frac{\rho}{\rho_{cp}}  \equiv \frac{N/V}{\sqrt{2}/D ^3}
\end{equation}
where $\rho_{cp}$, the number density at close packing, provides the natural scale.
The range of interest to us here extends from the melting density
$\tilde{\rho} \simeq 0.736$ \cite{meltingdensity} through to the close-packed 
limit $\tilde{\rho} =1$.

In the close-packed limit the configurational integral
(Eq.~\ref{eq:partconfwt}) may be
rewritten \cite{closepackedlimit} as the product of two terms:
\begin{equation} 
\label{eq:cppartconfwt}
\Omega (N, V, \alpha) = 
\Omega_0 (N, V)  \Omega_{\alpha}
\end{equation}
The first term here is defined by
\setcounter{abc}{1}
\begin{equation} 
\label{eq:omega0def}
\Omega_0 (N, V) =\left[ \frac{D \epsilon}{1-\epsilon} \right]^{3N}
\end{equation}
with
\addtocounter{equation}{-1}
\stepcounter{abc}
\begin{equation} 
\label{eq:epsilondef}
\epsilon \equiv 1- \tilde{\rho}^{1/3}
\end{equation}

The associated contribution to the entropy is logarithmically divergent 
in the close-packed limit \cite{classartifact}, but independent of the phase.
The second contribution to the configurational integral is defined by
\setcounter{abc}{1}
\begin{equation} 
\label{eq:omegaalphadef}
\Omega_{\alpha}
= \prod_i [ \int_{\framebox{$\alpha$}}  d\vec u _i ]\prod_{<ij>}^{nn}
\Theta (u_{ij}^{||} + 1 ) \left[1 +O(\epsilon)\right]
\end{equation}
where \cite{scaledvariables}
\addtocounter{equation}{-1}
\stepcounter{abc}
\begin{equation} 
\label{eq:uijdef}
\vec{u}_{ij} \equiv \vec{u}_i-\vec{u}_j \equiv 
u_{ij}^{||} \hat{n}^{\alpha}_{ij} +
\vec{u}_{ij}^{\perp}
\end{equation}
\setcounter{abc}{0}
while $\hat{n}^{\alpha}_{ij}$ is a unit vector from lattice site $j$ to 
nearest neighbor 
lattice site $i$.
The associated contribution to the entropy is finite, but
depends on the phase through the geometry of the nearest neighbor  vectors. It may be
visualized as that of a set of hard dodecahedra \cite{hardtosee}.

Now let us recall that the quantity of immediate
interest is the {\em difference} between the entropy-densities of the two phases. 
It may be written as
\begin{equation} 
\label{eq:entdiff}
\Delta s_{\alpha \beta} \equiv s(N,V,\alpha) - s(N,V,\beta) =\frac{1}{N} \ln
{\cal R}_{\alpha \beta} (N,V)
\end{equation}
where
\begin{equation} 
\label{eq:Rdef}
{\cal R}_{\alpha \beta} (N,V) \equiv 
\frac{\Omega(N,V,\alpha)}{\Omega(N,V,\beta)} 
=\frac{P(\alpha \mid N,V)}{P(\beta \mid N,V)}
\end{equation}
Here $P(\alpha \mid N,V)$ is the probability that a system, free to explore
the joint configuration space of the two structures (and visiting configurations
with the appropriate probabilities ---all equal in this case) will be found 
in some configuration of structure $\alpha$. 

In the constant density ensemble, then,  the computational task is
to determine the ratio defined by Eq.~(\ref{eq:Rdef}). In the constant pressure
ensemble we require the ratio ${\cal R}_{\alpha \beta} (N,P^{*})$
of the partition functions
\begin{equation}
\label{eq:constantPzdef}
{\cal Z}(N,P^{*},\alpha) = \int dV \Omega (N, V, \alpha) e^{-P^{*} V}
\end{equation}
where $P^{*}$ is a measure of the pressure \cite{units}.
The associated thermodynamic potential is the gibbs free energy density
defined by
\begin{equation}
\label{eq:partgibbs}
g(N,P^{*} ,\alpha) \equiv - \frac{1}{N} \ln {\cal Z}(N,P^{*},\alpha)
\end{equation}
so that, in analogy with Eq.~(\ref{eq:entdiff}),
\begin{equation} 
\label{eq:gdiff}
\Delta g_{\alpha \beta} \equiv g(N,P^{*},\alpha) - g(N,P^{*},\beta) =-\frac{1}{N} \ln
{\cal R}_{\alpha \beta} (N,P^{*})
\end{equation}

\subsection{The lattice-switch method \label{subsec:lsmethod}}

The two close-packed structures of interest here are shown schematically in
Fig.~(\ref{fig:stackpatt}). In principle there are many transformations 
which will map one set of lattice vectors into the other; we shall consider
the criteria guiding the choice in section~\ref{subsec:representations}.
The mapping used in most of the work reported here is
shown schematically in Fig.~(\ref{fig:latticeswitch}). 
This scheme exploits the
fact that the two structures differ only in respect of the stacking pattern of
the close-packed planes. A suitable transformation can then be constructed that
entails, simply, {\it translating} appropriate close-packed planes. By
`translate' we mean, more precisely, `relocate at a position defined by an
appropriate translation vector': one should not think of the planes as `sliding
through' the intermediate states.

Figure~(\ref{fig:latticeswitch}) shows the application only to the 
{\em perfect}-crystal-configurations where energy-matching is guaranteed
\cite{groundstatemismatch}.
In general (that is, for  `typical' configurations:
see Fig.~(\ref{fig:orderparam}) for an example) 
the two configurations related by the LS operation will not be
energy-matched: since adjacent planes are translated differently, the translations
may ---indeed, with overwhelming probability {\em will}---
map a realizable configuration (of one structure), in which there are no overlapping spheres,
onto an unrealizable configuration (of the other)
in which there are overlaps.
A  MC lattice switch `move' will be rejected for {\em most}
configurations. But not quite all: gateway configurations 
(configurations that are energy-matched \cite{energymatcheddef} 
to their conjugates)
must exist, in significant measure.
In particular, it is clear on grounds of continuity 
that  configurations `close enough' to
perfect-crystal form must fall into this category.
One might therefore {\em choose} these 
`small-displacement' configurations to act as the gateway states, and 
define a multicanonical weighting procedure accordingly. However,
one can avoid having to make this explicit choice,  and, instead, 
let the system {\em find} gateway configurations itself. 
To do so we must define a measure of the 
mismatch between the energies of the configurations linked by the
transformation. 

In the present context that mismatch is quantified by the
number of pairs of overlapping spheres created by the transformation.
To that end let $M(\{\vec{u}\},\alpha)$ denote the number of 
overlapping pairs associated with the displacements $\left\{\vec{u}\right\}$ 
within the structure $\alpha$. And define \cite{signconvention}
\begin{equation}
\label{eq:overlapdef}
{\cal M}(\{\vec{u}\}) 
\equiv 
M(\{\vec{u}\},{\mbox{\it hcp}}) 
-M(\{\vec{u}\},{\mbox{\it fcc}}) 
\end{equation} 
Since  $M(\{\vec{u}\},\alpha)$
will necessarily be zero for any realizable set of displacements of structure
$\alpha$, the {\em overlap order parameter} ${\cal M}$ is necessarily $\ge 0$
($\le 0$) for realizable configurations of the {\it fcc} ({\it hcp}) structure.
Figure~(\ref{fig:orderparam}) provides a concrete example. The gateway
configurations may then be identified ({\em without} prejudging their
microscopic character) as the set of configurations for which  ${\cal M}=0$:  a
displacement pattern $\left\{\vec{u}\right\}$ for which ${\cal M}=0$ is realizable in {\em both}
structures (energy-matched). A  LS MC step initiated
from an ${\cal M}=0$ configuration {\em will} be accepted; if initiated from outside this
set of configurations it will be rejected.

The sampling algorithm must thus be multicanonically customized so as to enhance
the probability along a notional line in  ${\cal M}$-space, extending from the
`equilibrium' ${\cal M}$-values (reflecting the number of overlaps created by a
LS acting on a {\em typical} configuration)  through to the ${\cal M}=0$ gateway
configurations.  This aim is realized by  augmenting the system  energy function
Eq.~(\ref{eq:hsinteraction}):
\begin{equation} 
\label{eq:wtinteraction}  E(\left\{\vec{r}\right\}) \rightarrow E(\left\{\vec{r}\right\}) + 
\eta ({\cal M}(\{\vec{u}\}) )  \equiv \tilde{E}(\left\{\vec{r}\right\})
\end{equation}
where $\eta ({\cal M}), {\cal M} =0,\pm 1,\pm 2 \ldots$ constitute  a
set of multicanonical weights \cite{bn}. 
These weights need to be chosen so as to allow the system
to access the ${\cal M}=0$ gateway
configurations, and thence (through the LS) the full joint configuration space of the
two structures.
The {\em desired} ratio of configurational weights, which reflects the {\em canonical}
distribution
\mbox{$P({\cal M}\mid N,V)$} 
(Eq.~\ref{eq:Rdef})
may then be estimated from the {\em measured multicanonical} distribution,
\mbox{$P({\cal M}\mid N,V,\{\eta ({\cal M})\})$} with the identification 
\begin{equation}
\label{eq:Rvalue}
{\cal R}_{fcc,hcp}(N,V)=
\frac{
\sum_{{\cal M} > 0} P({\cal M}\mid N,V)
}{
\sum_{
{\cal M} < 0} P({\cal M}\mid N,V)
}
=\frac{
\sum_{{\cal M} > 0} P({\cal M}\mid N,V,\{\eta ({\cal M})\})
e^{ \eta ({\cal M})}
}{
\sum_{
{\cal M} < 0} P({\cal M}\mid N,V,\{\eta ({\cal M})\})
e^{ \eta ({\cal M})}
}
\end{equation} 
Here the exponential re-weighting of the multicanonical distribution folds out
the bias associated with the weights, whose residual effects are then simply
as desired ---the
removal of the ergodic barrier between the two branches of the distribution.

\subsection{Representations: tuning the lattice switch
\label{subsec:representations}}

We have presented the LS method in its simplest realization  --the
one we have used for most of the studies reported here. 
We now outline two important respects  \cite{notreallyseparate} 
in which some degree of generalization is possible, and may be desirable,
in subsequent applications. Both involve the choice of {\em representation} of the LS transformation.

We have already alluded to the first point: there are many forms of
lattice-to-lattice mapping. It is clear that the efficiency of the  method  will
depend significantly upon the mapping chosen. Evidently the choice should be
made so as to match up, as closely as possible, the {\em energy} of the two configurations it links.  In
the  context of hard spheres this aim is realized by choosing the mapping which
gives the smallest  equilibrium overlap-count (mean $\mid {\cal M} \mid$-value), which gives a measure of 
the entropic barrier that has to be negotiated  by the multicanonical
procedure. The smaller this barrier, the shorter is the path to the gateway
configurations from which a successful LS may be launched. Since the multicanonical
simulations traverse this path only slowly (essentially diffusively, at best)
the gains here are potentially substantial.
It is intuitively clear that the scheme described above will fulfill  this
criterion well: in this representation, the LS translates close-packed planes 
bodily,  so  it can create overlaps only between spheres associated with {\em
different} planes.  But it is useful to explore other schemes --partly  to check
that there is no significantly better alternative, but principally to understand
the different  factors that control the efficiency. We have done so; the results
are to be  found in section~\ref{subsec:effofrep}.

There is a second 
 -- less obvious-- generalization of the framework.
In the simple realization, the particle positions are written
in the `lattice plus displacement'  representation provided
by Eq.~(\ref{eq:decomposition}). The LS operation then maps a configuration 
of one structure onto a configuration of the other with the {\em same} set of
displacements. This is unnecessarily restrictive. More generally 
we are at liberty to write, in place of Eq.~(\ref{eq:decomposition}),
\begin{equation}
\label{eq:gendecomposition}
\vec{r}=\vec{R}^{\alpha}+{\bf T}^{\alpha} \cdot \vec{u}
\end{equation}
where $\vec{r}$, $\vec{R}^{\alpha}$ and $\vec{u}$ are now
column vectors with $3N$ elements and ${\bf T}^{\alpha}$ is a $3N \times 3N$
non-singular matrix, whose form (possibly $\{\vec{u}\}$-dependent) is  at our disposal. 
Eq.~(\ref{eq:partconfwt}) is then replaced by
\begin{equation} 
\label{eq:genpartconfwt}
\Omega (N, V, \alpha) = \prod_i [ \int_{\framebox{$\alpha$}}  d\vec u _i ] 
\cdot  \det {\bf T}^{\alpha}\prod_{<ij>}\Theta (r_{ij} -D )
\end{equation}
From the standpoint of the (standard) single-phase part of the MC procedure, this
change in representation is equivalent to  changing the form of
the configurational energy:
\begin{equation} 
\label{eq:detinteraction}  E(\left\{\vec{r}\right\}) \rightarrow E(\left\{\vec{r}\right\}) 
-\ln \left[ \det {\bf T}^{\alpha} \right]
\end{equation}
This change introduces some computational overheads, which could be substantial
if the {\bf T}-transformation is not local. The potential pay-off
lies in the LS part of the MC procedure. One might hope to be able to tune the
form of the  {\bf T}-matrix so that `typical' configurations of the one 
structure are mapped (by LS) into `typical' configurations of the other.
In the case of the hard sphere problem, however, our results
(section~\ref{subsec:effofrep}) suggest that there is little to be gained
here by this kind of tuning.

\section{Implementation \label{sec:implementation}}

\subsection{Monte Carlo procedures\label{subsec:MCprocedures}}

First we consider the procedure for MC-sampling of the particle displacements,
for a given structure (set of lattice vectors). As discussed in
section~\ref{subsec:model} this sampling should, in principle, satisfy some
appropriate configurational constraint \cite{ifitsnotthere}. In our original
studies \cite{lsprl} we chose to implement this constraint {\em explicitly},
through our sampling distribution: candidate displacements were drawn from a
flat (`top-hat') distribution. This procedure can be made
to work.  But the constraint explicitly breaks the translational invariance; and
one must deal with the consequences. In particular the configurational integral
effectively being evaluated  then {\em depends} upon the location of the center
of mass and thence upon the top-hat cut-off; this dependence sets in when the
displacement acquired by the center of mass,  in the course of its slow diffusive motion, becomes
comparable with the top-hat-cut-off. One can avoid this problem simply by 
fixing the
center of mass. Our failure to do so in \cite{lsprl} led to results which differ
significantly from those we present here. In the studies reported here
we have chosen the `implicit' realization of the configurational constraint
(practically, but not conceptually  equivalent to ignoring it) which rests
(section~\ref{subsec:model}) on time scales. 
Spheres were chosen at random, and trial-{\em changes} to the current
displacement  drawn from a uniform distribution. 
The displacement update is accepted according to the Metropolis prescription
\cite{binrev}
\begin{equation} 
\label{eq:uupdateacc}
p_a (\left\{\vec{u}\right\} \rightarrow \left\{ \vec{u}^{\prime}\right\} )
= \mbox{min}\left\{1,\exp\left[-\Delta \tilde{E}(\left\{\vec{r}\right\})\right] \right\}
\end{equation}
where $\tilde{E}(\left\{\vec{r}\right\})$ is defined in Eq.~(\ref{eq:wtinteraction}). 
In addition to the 
constraint that the update should yield a realizable configuration of the 
current phase, this acceptance probability reflects the chosen weights
which are defined (section~\ref{subsec:weights} explains how) on the space of the overlap order parameter ${\cal M}$ 
(Eq.~\ref{eq:overlapdef}).
To minimize the computational time spent determining how 
a proposed move affects the value
of $\mathcal{M}$ we used a local overlap array, holding
information on which neighbors of a given sphere
currently overlap with that
sphere  in the conjugate
configuration 
generated by a LS.

The representation of the close-packed limit provided by 
Eq.~(\ref{eq:omegaalphadef}) can be handled with only minor amendments: 
the constraint $r_{ij}>D$ identifying realizable 
configurations is replaced by a constraint on the scaled displacement-difference coordinates,
$u_{ij}^{||}>-1$. The overlap order parameter (measuring the number of times the hard
sphere constraint is violated in the conjugate configuration) is
redefined accordingly. 
In this limit particle `interactions' (encounters) may occur only between
immediate neighbors. At other densities we allowed for the 
possibility of encounters between nominal second-neighbors. 
We found however that although the number of such encounters grows
rapidly with the approach to the melting density, the consequences for the 
relative entropy of the two structures is insignificant under the conditions
studied here \cite{sortofmauhuse}.

In addition to particle moves the constant-pressure simulations require 
updates of the simulation-cell-parameters. 
In such an update (implemented on average once
per sweep) a trial set of cell-parameters
are selected, and accepted with probability \cite{vupdateref}
\begin{equation} 
\label{eq:vupdateacc}
p_a (V \rightarrow V^{\prime} )
= \mbox{min}\left\{1,\exp\left[
-\Delta \tilde{E}(\left\{\vec{r}\right\})  -P^{*} \Delta V + N\ln(V^{\prime}/V)\right]\right\}
\end{equation}
where $V'$ is the volume associated with the trial parameters.
Note that this kind of  update --a dilation--  changes $\tilde{E}(\left\{\vec{r}\right\})$ both trivially
(so as to forbid moves causing `real' overlaps) and more subtly through changes
in the count of the overlaps in the conjugate configuration. A volume update
thus requires recalculation of the entire
local overlap array.

Now consider the lattice switch. The switch may be viewed as an updating of
the `lattice'-type $\alpha$, regarded as a stochastic variable. The prescription
for such an update is quite simple. 
After every particle update the value of ${\cal M}$ is checked 
(it is already known).  The LS is performed if
(and only if) the gateway condition ${\cal M} = 0$ is satisfied.

\subsection{Calculating the weights \label{subsec:weights}}

The determination of an acceptable set of multicanonical weights \cite{bn}
can be accomplished in a number of ways 
---none, seemingly, entirely systematic. 
We describe briefly the techniques we have used in the present study. Figure
~(\ref{fig:weightgeneration}) 
provides some illustration. For
further details and references  to other work the reader is referred to 
\cite{bn,grsadbspt,grsadbmultican,fitzgerald}.

The simplest method is the {\it{visited states}} (VS) technique
\cite{grsadbmultican}. In this approach a suitable set of weights is evolved through
an iterative process (Fig.~\ref{fig:weightgeneration}),  the next set of weights
depending upon the distribution of the (overlap) order parameter over the
macrostates visited using the current set of weights. This process is repeated
until  the weights yield a  distribution $P({\cal M}\mid N,V,\{\eta ({\cal M})\})$
that is effectively flat. This method proved quite adequate for our smallest
system.

For larger systems, however, we found it more efficient
to appeal to the {\it{transition probability}} (TP) method
\cite{grsadbmultican}. In the simplest realization of this method
the simulation is initiated from a `cold' (zero-displacement)
configuration (a member of the ${\cal M} = 0$ macrostate) for one structure. In the course of its
subsequent evolution towards equilibrium for that structure the numbers of
transitions between different ${\cal M}$-macrostates 
are recorded, and subsequently used to construct an estimator of
the macrostate-transition-probability matrix. This TP matrix 
can be used to estimate the macrostate probability distribution and thence
to provide an estimate for a set of weights (Fig.~\ref{fig:weightgeneration}), which can
in turn be refined via VS. For our intermediate size system
this method worked well.

In the case of our largest system we found it necessary to modify the method
somewhat, so as to limit the rate at which  the simulation passes through
${\cal M}$-space. One way of doing this is to constrain the system to macrostates with
overlap order parameters  below some `barrier'-value, which is gradually
incremented (moved `out' in ${\cal M}$-space),  at intervals  of the order of the
equilibration time.

By {\it fiat} the two structures have the same weights
for ${\cal M} =0$.
In principle, the  weights associated with the two 
structures for non-zero $\mid{\cal M}\mid$ are different 
(i.e. $\eta({\cal M}) \neq \eta({-\cal M})$), 
and have to be evolved separately.
In practice
the weights of the two structures are
very similar ---a reflection of the similarity of the 
entropies of the two phases.
Consequently,
one set of weights provides an excellent first approximation 
 to the  other, for refinement by VS.

\subsection{Simulation details}

The specific form of the LS operation we have chosen
(Fig.~\ref{fig:latticeswitch}) imposes restrictions on the geometry of the
system simulated: with normal periodic boundary conditions the system must
comprise integral multiples of  6 close-packed planes. It is possible to avoid
this restriction by using more elaborate boundary conditions \cite{mauhuse}, but
we chose to avoid this complication and simulate systems comprising $6^3=216$,
$12^3=1728$ and $18^3=5832$ spheres. 
Simulations were performed at two densities, namely (cf Eq.~\ref{eq:densitydef}) 
$\tilde{\rho}= 0.7778$ \cite{whythisdensity}, and 
$\tilde{\rho}= 1$, the close-packed limit.

The maximum step size for displacement updating  was chosen so as to minimize
the autocorrelation time of the overlap order parameter
(Eq.~\ref{eq:overlapdef}). We found a maximum step size of $0.13D$ produced
the best results at  $\tilde{\rho}= 0.7778$, while a value close to unity
was found to be appropriate in
the close-packed limit,  in the representation (and scaled units)
given in Eq.~(\ref{eq:omegaalphadef}).

A significant proportion of our simulation time was devoted to the process of
weight-determination. For our largest system we used $10^6$ Monte Carlo sweeps
(MCS) to generate a first (TP) estimate of the weights, with a further $5 \times
10^6$ MCS devoted to weight-refinement using VS. 

The  free-energy differences of interest were then determined 
by further simulations in the multicanonically-weighted ensemble.
For each system (density, and size) we performed a series of
runs each long  on the scale of the
autocorrelation time of the overlap order parameter. Each of these runs
then provides an independent estimate of the (logarithm of the)
probability ratio required (Eq.\ref{eq:Rvalue}).
The standard deviation of these estimates provides a basis for assigning
an associated statistical uncertainty. Implementing this stage
required  simulation times ranging from 
$\sim 2.5 \times 10^8$ MCS for $N=216$ to
$\sim 4 \times 10 ^7$ MCS for $N=5832$. 

\section{Results \label{sec:results}}

\subsection{The effects of the representation \label{subsec:effofrep}}

As discussed in section~\ref{subsec:representations} the LS operation
can be implemented with different choices of representation of 
the lattice-mapping or the particle displacements \cite{notreallyseparate}.  

The efficiency of a  lattice mapping is measured (inversely) by the equilibrium
overlap count. Table~\ref{tab:switch} shows results for a variety of mappings,
chosen to expose the different factors that control the mapping efficiency.
Mapping number 1 is the one described in
Fig.~\ref{fig:latticeswitch}, and used throughout this work: the 
notation $(0, -\vec{t}, +\vec{t})$ signifies that the three pairs of planes
counting from the top of  Fig.~\protect(\ref{fig:latticeswitch})
are translated respectively by $0$, $-\vec{t}$ and $ +\vec{t}$.
A similar convention is used to label mappings 2 and 3. 
In mapping 4 (`random-plane') an {\it hcp} configuration is generated by taking
an {\it fcc} configuration and restacking  its close-packed planes in a random
order, in an {\it hcp} pattern.
In mapping 5 (`random-site') an {\it hcp} configuration is generated by mapping
the particle displacements in an {\it fcc} configuration randomly on to the
sites of an {\it hcp} lattice.

The random-site mapping (number 5) shows the largest overlap count. One can account for
its value, rather well, by regarding the particle displacements  as {\em
isotropic}, {\em gaussian}  and {\em independent of structure}
\cite{displacements}, and estimating the probability that two particles
associated with nearest-neighbor sites, and with displacements drawn {\em randomly} from this
distribution, will overlap.

Using the random-plane mapping (number 4) cuts the overlap count by a factor of
(a little more than) two with respect to random-site.  This efficiency gain simply
reflects the fact that of the $6N$ potential overlaps between near-neighbors,
only the $3N$ associated with neighbors in different (but adjacent) planes can
now contribute. 

Mapping 3 simply generates one {\it fcc} configuration from another (it is
useful only because it is informative): its overlap count is cut by a further
factor of two. This reflects the fact that  this mapping (like mappings 1 and 2)
moves close-packed planes in {\em pairs}, thus guaranteeing no overlaps between
the two members of each pair.

Mappings 2 and 1 show further ---smaller but still practically useful--- cuts in
the overlap count. The origin of these gains is more interesting. It is clear
that they must reflect the size of  the translation vector used: mappings 1
through 3 differ only in this respect. This vector  controls the extent of the
shear which the mapping  introduces between successive pairs of planes.  The
following interpretation seems  reasonable. The displacement patterns in
adjacent planes will be correlated to some extent, with undulations in one
surface (the z-components of the displacements) matched to undulations in its
neighbor.   The smaller the shear, the  more closely  these undulations will
{\em remain} matched to one another (in the conjugate configuration), and the
smaller the overlap count. With increasing shear, this advantage is lost and the
behavior should (and indeed does) approach the limit (one quarter of the overlap count for mapping 5) one would expect in the absence of such correlations.
The fact that this `approach' is already apparent in the performance
of mapping 2 is consistent with the fact that  the measured correlation length
of the surface undulations   at the density concerned is found to be
close to  the magnitude  of the translation vector $\vec{t}$.

These results help to clarify the factors which control the overlap
count of the mapping (number 1) we have actually used. 
It is tempting to attribute the overlaps to the fact that the LS ({\it fcc} $\rightarrow$
{\it hcp}, say)
maps each particle from an environment in which adjacent close-packed planes 
have different stacking labels (A and C, say) to one in which they have the
same label (C, say). The results for mappings 1-3 show that it would be 
misleading to think this way. The overlaps
simply reflect the numbers of particles that `see' a new adjacent close packed plane (irrespective of its label), and the extent to which it is `new'.
This is the reason for the  similarity
between the overlap counts for the two 
structures (section~\ref{subsec:results}).
It shows, moreover, that any simple \cite{whatissimple}
tuning of the displacement representation
(the choice of $T$-matrix) is likely to be of no advantage here
\cite{fantasyland}.

\subsection{How it works: the gateway configurations
\label{subsec:gatewaystates}}

A LS operation will work (be accepted as a MC move) only when launched from a
small subset of the configurations actually visited: these, by definition, are
the `gateway configurations'. As noted earlier, one could identify {\em a priori}
configurations (those characterized by `small enough' displacements) which fall
into this set. But we have elected, rather,  to let the system (the algorithm)
identify them on the basis of their defining characteristic --- that they have
zero overlap order parameter  ${\cal M}$ \cite{gatewaystatedef}.
It is then interesting to investigate
the {\em microscopic} characteristics of the configurations picked out by this
constraint. Figure~(\ref{fig:sepdist}) shows the distribution of the 
separation, $d$, between adjacent close-packed ($x-y$) planes \cite{moreprecisely},
for  ${\cal M}$-macrostates
corresponding to equilibrium {\it fcc},  equilibrium {\it hcp},
and gateway (${\cal M}=0)$ regions. 
The macrostates corresponding to the equilibrium 
crystal structures have similar, near-gaussian, $d$-distributions. In contrast,
for the gateway macrostate the distribution is {\em bi-modal}: 
in this macrostate,
some planes are systematically moved closer to one  another,
while (in equal measure) others are shifted apart. On closer examination one
finds that it is the planes which are {\em translated
together} by the LS  (eg the pair of planes marked (i) in
Fig.~(\ref{fig:latticeswitch})) that fall into the first category, while
the planes that are {\em translated
differently} by the LS  (eg the pair of planes marked (ii) in
Fig.~(\ref{fig:latticeswitch})) fall into the second. The evolution, with ${\cal M}$ , 
of the mean plane separation (for both categories) is shown in
Fig.~(\ref{fig:gatewayfeatures}a).  
The behavior thus unearthed is entirely reasonable.
The LS operation can {\em only} create overlaps between
neighboring planes which are translated by different amounts (sheared with
respect to one another). The algorithm resolves the task set by the bias towards 
${\cal M}=0$ by moving these pairs of planes (the ones vulnerable to overlaps)
further apart, at the expense of a compression of the others
\cite{morecleverstill}. 
In simulations conducted at constant pressure this effect (still present) is
supported by a second.  Fig.~(\ref{fig:gatewayfeatures}b) shows that the
algorithm now exploits the additional degrees of freedom (the {\em shape} of
the simulation cell) to locate  gateway states with  values of the $c/a$ ratio
enhanced above the ideal close-packed value \cite{ca-ratio}. Again, the
advantages with respect to the switch are clear.

It is tempting to say that the sampling is intelligent.
In any event
it is clear that the algorithm locates and utilizes configurations
which it would be difficult to exploit explicitly in the design
of the switch operation.

\subsection{Entropies of crystalline structures \label{subsec:results}}

The essential output of a LS-simulation is in the form of the normalized
probability distribution of
the overlap-order-parameter, reweighted to remove the  bias in the
multicanonically-weighted  distribution actually measured.
Figure~(\ref{fig:pvsmvarious}) shows the results for this distribution
(at 
$\tilde{\rho}=0.7778$) for three different $N$ values. As one would expect the
distributions each comprise two  peaks (one associated with each  phase)
each of which is {\em nearly} gaussian \cite{butnotquite}
and sharpens with increasing $N$ \cite{whenviewed}. Note the close
correspondence between the equilibrium overlap counts for
the two structures. This result is not {\em required} by definition, or any
obvious symmetry. Rather it should be seen as a further manifestation (the
smallness of the entropy difference between the phases is the prime one) of the
similarity of the local particle environments in the two structures.

The relative weights of the two peaks is a direct measure
of the difference between the entropies of the two structures
(Eqs.~\ref{eq:entdiff}, \ref{eq:Rdef}, \ref{eq:Rvalue}). 
Since the entropies are extensive the ratio of the peak weights
grows exponentially with $N$ \cite{notaproblem};
the fact that (in this case, at least for our smaller systems) 
the two peaks can even be displayed on the same scale
is a reflection of the exceptionally delicate balance between the two entropy densities.

Figure (\ref{fig:pvsmvarious}) allows one to {\em see} that {\it fcc} is the 
thermodynamically preferred structure. This conclusion is expressed
quantitatively in the results gathered in Table~\ref{tab:sdiff}.  Our results
at $\tilde{\rho}=0.7778$ correct those of our earlier work  \cite{lsprl}, as
explained in section~\ref{subsec:MCprocedures}. They are in full accord with the
results (both LS and IM-based)  reported by Pronk and Frenkel \cite{pronkfrenkel}. The close
correspondence between  the results for $N=1728$ and $N=5132$ confirms that the
former system is already representative of the thermodynamic limit. 
Table~\ref{tab:sdiff} also shows the results of our studies at the
close-packed limit, using the hard-dodecahedron
representation (Eq.~\ref{eq:omegaalphadef}). 
Our results seem at variance with
the IM-based result of Woodcock \cite{woodcockcorr}, even allowing for
the large uncertainty attached to that result. They are close to
those  (based on LS) reported by Mau and Huse \cite{mauhuse}. 
But the differences (for the smaller systems, particularly) appear to be
statistically significant \cite{checkofclosepackedlimit}.
Figure~(\ref{fig:speedycurve}) gives an alternative view of these results.
It utilizes the parameterization of the measured pressure 
difference between the two phases provided by Speedy \cite{speedy} 
to determine the entropy difference, as a function of density, {\em given} the
entropy difference at a chosen reference density; we have used the results of
the present work at 
$\tilde{\rho}=0.7778$. 

Table~\ref{tab:gdiff}  shows the results of our studies in the constant
pressure ensemble. The quantity of interest here is the difference between the
gibbs free energy densities at the chosen pressure, which follows from the
relevant distribution  with the aid of Eq.~(\ref{eq:gdiff}). In fact the gibbs
free energy density difference $\Delta g$ for a given pressure, and the entropy density
difference $\Delta s$ at a physical density that is the thermodynamic conjugate of that
pressure for one of the phases, differ (in magnitude \cite{butsign}) by terms that are {\em second order} in
the pressure difference between the two phases.  
That pressure difference is extremely small \cite{speedy}, as is the 
difference between the measured densities of the two structures 
(Table~\ref{tab:gdiff}).
In these circumstances one would expect the magnitude of 
$\Delta g$  to fall on the  $\Delta s$ plot 
in Figure~(\ref{fig:speedycurve}); and indeed, within the residual uncertainties,
it does.

\section{Discussion: review and prospects \label{sec:prospects} }

In the work described here we have been concerned both with a {\em system} of
long-standing interest -the hard sphere crystal-- and a  {\em method} --lattice-switch Monte Carlo-- with potentially wide applicability. We divide our
concluding discussion accordingly.

The full agreement between the present work and that of \cite{pronkfrenkel}
leaves little doubt that the equilibrium entropy difference between the two
close-packed structures has finally been established securely and with high
precision ---at least at one density. Although a small discrepancy with respect 
to the results of \cite{mauhuse} remains, the accord  of our close-packed  limit
results with those established using pressure difference measurements
\cite{speedy} suggests that the curve in Fig.~(\ref{fig:speedycurve}) provides a 
relatively complete and trustworthy picture of the density-dependence.

Notwithstanding the simplicity of the model, these results do have implications
for experimentally-realizable systems.  The immediate  relevance to atomic
systems is tenuous \cite{choi}, but the model has been widely used to account
for the behavior of assemblies of  `hard', `spherical' colloidal particles \cite{pnphsrev}.
Since the predicted entropy-density difference is so small
there are potentially many ways (residual interactions between the spheres;
polydispersity) in which the applicability of the theory may be
compromised. But, of these, it seems that the most significant issues to be
addressed are to do with {\em scales} --length and time.

First, the lengths. In the experiments reported in \cite{pnphsprl} the colloidal
particles have diameters of order $10^{-7}\,m$ and the samples
comprise crystallites with linear dimensions of order $10^{-5}\,m$.  The 
number of particles in such a crystallite ($N\sim 10^6$) is large compared to
those in our simulation, which is (as we have seen) sufficient to allow us to
{\em deduce} properties of the thermodynamic limit.  But it is {\em not} large enough 
to guarantee that the behavior displayed will actually {\em be}
that of the thermodynamic limit. To see this --and its principal implications-
one needs to consider the stability of the perfect {\it fcc} crystal with respect to
{\it hcp}-type stacking faults. Following reference \cite{pnphsprl} we may
introduce a parameter $\alpha$ \cite{notstackinglabel} measuring the probability that a 
chosen close-packed  plane sits within an {\it fcc} environment
as distinct from the {\it hcp}
environment.
A simple argument (Appendix A) using the
pseudo-spin parameterization of stacking patterns provided in \cite{mauhuse}
then yields the result 
\begin{equation}
\label{eq:alpha}
\alpha = \frac{1}{2} \left( 1+\tanh \left [\frac{N_{\perp} \Delta s}{2} \right] \right )
\end{equation}
where $N_{\perp}$ is the number of particles in a close packed layer and
$\Delta s$ (a function of $\tilde{\rho}$) is the {\it fcc-hcp} entropy difference per particle, as given in
(and in the units of) Table~\ref{tab:sdiff}. The thermodynamic ideal
($\alpha=1$)
is thus realized only to the extent that $N_{\perp} \Delta s$ is large
compared to unity. For the length scales given above,
$N_{\perp} \Delta s \simeq 1$. The obvious implications are qualitatively consistent
with the observations reported in 
\cite{pnphsprl} which show $\alpha$ values (deduced from  Bragg
scattering intensities) ranging from $0.5$ (signaling essentially
random-hexagonal-close packing, {\it rhcp}) through to $\alpha = 0.8$.

The observed spread in $\alpha$ values reflects --presumably-- the issue of time
scales. The smallness of the entropy difference  (which supplies the kinetic
driving force towards the equilibrium state) suggests that the equilibrium
behavior will be observed only in samples which are grown sufficiently slowly
and (or) given sufficient time for subsequent annealing \cite{frenkelsestimate}.
The results of \cite{pnphsprl} do indeed suggest a correlation between observed $\alpha$  value
and the slowness of the growth process. Experiments done in microgravity
\cite{spaceshuttlehs}, where growth processes  are greatly accelerated, yield
essentially randomly close-packed crystals.

Now let us turn to the lattice-switch method. There are two questions here:
One: does the method represent a significant advance with respect to
existing methods? Two: is it generally applicable?

The main alternative method (the benchmark against which others need to be
assessed) is probably integration along a reference path, of which the work
reported in \cite{pronkfrenkel} represents, to our knowledge, the most refined
example.  If one compares the two techniques (LS and IM) on the basis of
precision-for-computational-buck there seems to be no clear winner in the
hard-sphere studies to date: reference \cite{pronkfrenkel} reports calculations
using both methods that achieve comparable levels of precision on the basis of
comparable computational time. But one should note
that the entropy
difference ultimately determined is some {\em four orders of magnitude} smaller
\cite{seeoldfrenkelladd} than the separate entropies of the two phases,
determined by IM. One can see this as a testimony to the care with which the
recent IM studies  have been carried out; or (as suggested in
section~\ref{SEC:INTRO}) as  a strong indicator that another
approach using an inter-phase path
is called for. There are also two other counts ---both somewhat subjective--- on
which we suggest that the LS approach is superior. First, it seems to us
relatively illuminating (by comparison with IM)   to read-off the result for a
free energy difference directly from a figure like Fig.~(\ref{fig:pvsmvarious})
which {\em shows}  what it {\em means}. Secondly  it also seems to us that LS
wins in regard to the transparency of the uncertainties to be attached to its
results. The LS error bounds represent  purely statistical uncertainties
associated with the measurement of the relative weights of two
distribution-peaks. The IM error bounds have to aggregate the uncertainties
associated with different stages of the integration process.

As regards the second question, we expect that the method will,  with
appropriate extensions, be widely applicable.  The first extension must clearly
be to accommodate soft potentials. The LS operation will then  need gateway
configurations in which the energies of the two structures (measured with respect to
their ground-state energies ---or indeed {\em any} fixed reference energy) are
closely-matched  \cite{groundstatemismatch}. The `overlap order parameter' will
need to be redefined accordingly. With no more than this degree of elaboration
the method should be applicable immediately to investigate the widespread
`competition' between {\it fcc} and {\it hcp} ordering in the phase-behavior of
the elements \cite{phasesofelements}. 

More generally, moving beyond  the space of {\it fcc-hcp}  structures, 
the choice of lattice-to-lattice mapping will require some thought. Mappings
which  preserve the relative positions of significant subsets of the particles
(the analogues of the close-packed planes) are likely to be be optimal.
The license to choose ones
representation of the displacements 
(Section~\ref{subsec:representations}) may also prove useful. Simple 
transformations \cite{whatissimple} will help if the mapping takes particles
between environments in which the  spectrum of single-particle displacements
is significantly different. In such cases one might envisage using a MC-annealing procedure to 
refine the choice of representation.  
The use of normal coordinates has some advantages here --but possibly not enough to
offset the fact that the interaction potential is non-local when expressed in fourier
space.

\acknowledgements

ANJ acknowledges the support of the EPSRC
through a research studentship.
NBW acknowledges the financial support of the Royal Society (grant no.
19076), the EPSRC (grant no. GR/L91412) and the Royal Society of
Edinburgh. ADB acknowledges helpful discussions with Dr Mike Evans.

\appendix

\renewcommand{\theequation}{\Alph{section}\arabic{equation}}

\section{Displacement entropy versus stacking entropy}

Consider a system of $N$ hard spheres arranged in $N_{\parallel }$ close-packed
layers of $N_{\perp}$ particles. Following reference \cite{mauhuse} one may
conveniently index each of the close-packed layers with a pseudo-spin
(Ising-like) variable $\sigma$, where $\sigma_i =1$ signifies
that layer $i$ has an {\it fcc} environment (the two immediately adjacent layers are
not aligned with
one another) while $\sigma_i =-1$ implies an 
{\it hcp} environment (adjacent planes are aligned with one another). 
The probability of a particular stacking sequence $\{\sigma\}$
(if these variables may properly be regarded as annealed) then satisfies
\begin{equation} 
\label{eq:stackprob}
\ln P(\{\sigma\}|N,V) = S(N,V, \{\sigma \}) +constant
\end{equation}
where $S(N,V, \{\sigma \})$ measures the entropy associated with the
configurations (displacements) consistent with the particular structure $\{\sigma \}$.
Following \cite{mauhuse} this entropy 
(we will refer to it here as `displacement entropy')
can usefully be written in the form of an expansion:
\begin{equation} 
\label{eq:dispentexp}
S(N,V, \{\sigma \}) = Ns_0 
+ N_{\perp} h \sum _i \sigma _i
+ N_{\perp} J \sum _{<ij>} \sigma_ i \sigma_ j + \ldots
\end{equation}
The expansion is effectively ordered in the {\em range} of the
entropic inter-layer `interactions': the ellipsis represents 
contributions from interactions (microscopically, displacement-displacement
correlation functions) extending over more than 4 layers. The analysis of
reference \cite{mauhuse} indicates that the series converges quickly, except
close to melting. If we neglect the interaction terms altogether 
we may make the identification
\begin{equation} 
\label{eq:efffield}
h= \frac{1}{N_{\perp}N_{\parallel }}\left[S(N,V, \{\sigma =+1\}) - S(N,V, \{\sigma
=-1\}) \right] = \frac{\Delta s_{fcc,hcp}}{2} 
\end{equation}
and, from Eq.~(\ref{eq:stackprob}), 
\begin{equation}
\label{eq:avsigma}
\langle \sigma \rangle = 
\frac{1}{N}\sum _{\{\sigma\}, i}
P(\{\sigma\}|N,V)\sigma_i=\tanh \left[N_{\perp}h \right] =  
\tanh \left[\frac{N_{\perp}\Delta s_{fcc,hcp} }{2} \right]
\end{equation}
from which Eq.~(\ref{eq:alpha}) follows. The correspondence with a 1D paramagnet
is clear. The familiar competition (between orientation energy and entropy) 
is played out here as a competition between displacement entropy and stacking
entropy, with $N_{\perp}$ playing the role of an inverse temperature.

\newpage

\begin{table}[htb!]

    \begin{center}
    \begin{tabular}[h]{cccc} 
      mapping  &description& effect & $m={\cal M}/N$\\
      \hline
1& $(0, -\vec{t}, +\vec{t})$  & {\it fcc} $\rightarrow$ {\it hcp} &  0.150(1) \\
2& $(0, 2\vec{t}, -2\vec{t})$ & {\it fcc} $\rightarrow$ {\it hcp} &  0.183(1) \\
3& $(0, 3\vec{t}, -3\vec{t})$ & {\it fcc} $\rightarrow$ {\it fcc} &  0.194(1) \\
4& random-plane               & {\it fcc} $\rightarrow$ {\it hcp} &  0.373(2) \\
5& random-site                & {\it fcc} $\rightarrow$ {\it hcp} &  0.820(3) \\
    \end{tabular}
    \end{center}

\caption{The efficiency of different lattice mappings (for $N=1728$ and 
$\tilde{\rho} = 0.7778$), as measured by the number
of overlaps (per sphere) that they generate.  Refer to the text for details.
\label{tab:switch}
}
\end{table}

\newpage

\begin{table}[htb!]

    \begin{center}
    \begin{tabular}[h]{rrrrrr}
      $\rho / \rho_{cp}$ & N & \multicolumn{2}{c}{$\Delta s$  $(10^{-5}\times k_B )$} & Method & Ref.\\
      \hline
      0.731  &   512 &   85 &  (10) &    SM &\cite{mauhuse}\\
      0.736  & 12000 &  230 & (100) &   IM &\cite{woodcockcorr}\\
      0.736  & 12096 &   87 &  (20) &    IM &\cite{bolfrenkelmh}\\
      0.739  &   512 &   90 &   (4) &    LS &\cite{mauhuse}\\
      0.7778 &   216 &  132 &   (4) &    LS &\cite{pronkfrenkel}\\
      0.7778 &  1728 &  112 &   (4) &    LS &\cite{pronkfrenkel}\\
      0.7778 &  1728 &  113 &   (4) &    IM &\cite{pronkfrenkel}\\
      0.7778 &   216 &  133 &   (3) &    LS &PW\\
      0.7778 &  1728 &  113 &   (3) &    LS &PW\\
      0.7778 &  5832 &  110 &   (3) &    LS &PW\\
      1.00   & 12000 &  260 & (100) &    IM &\cite{woodcockcorr}\\
      1.0    &   512 &  110 &  (20) &    SM &\cite{mauhuse}\\
      1.0    &    64 &   91 &  (5)  &    LS &\cite{mauhuse}\\
      1.0    &   216 &  107 &  (4)  &    LS &\cite{mauhuse}\\
      1.0    &   512 &  119 &  (3)  &    LS &\cite{mauhuse}\\
      1.0    &  1000 &  113 &  (4)  &    LS &\cite{mauhuse}\\
      1.0    &   216 &  131 &   (3) &    LS &PW\\
      1.0    &  1728 &  125 &   (3) &    LS &PW
    \end{tabular}
    \end{center}

\caption{
\label{tab:sdiff}
The difference in the entropy
densities 
of the $fcc$ and $hcp$ structures,
$\Delta s\equiv \Delta s_{\mbox{\it fcc,hcp}}$ (Eq.~\protect\ref{eq:entdiff});
the  associated uncertainties are in parenthesis.
The results of the present work (PW)  supercede those of reference~\protect\cite{lsprl}.
The results of reference \protect\cite{woodcockcorr} supercede those of
reference \protect\cite{woodcock}. IM stands for integration method; SM
is  the lattice shear method of \protect\cite{mauhuse,shearmethod}.
}
\end{table}

\newpage

\begin{table}[htb!]

    \begin{center}
    \begin{tabular}[h]{cccccc} 
      $P^{\star}$ $ (D^{-3})$ & $\tilde{\rho}_{hcp}$ & $\tilde{\rho}_{fcc}$ & N & \multicolumn{2}{c}{$\Delta g$ $(10^{-5}\times k_BT )$}\\
      \hline
      14.58 & 0.7776(1) & 0.7775(1) & 216    &  -113 & (4) \\
      14.58 & 0.7770(3) & 0.7774(2) & 1728   &  -115 & (6) \\
    \end{tabular}
    \end{center}

\caption{
\label{tab:gdiff}
The difference in the gibbs free energy densities
of the $fcc$ and $hcp$ structures
$\Delta g \equiv \Delta g_{\mbox{\it fcc,hcp}}$ (Eq.~\protect\ref{eq:gdiff});
the  associated uncertainties are in parenthesis. $P^{\star}$ gives the 
pressure \protect\cite{units} in units  of $k_BT/D^3$. 
}
\end{table}

\newpage

\begin{figure}[h]
\epsfxsize=150mm
\epsffile{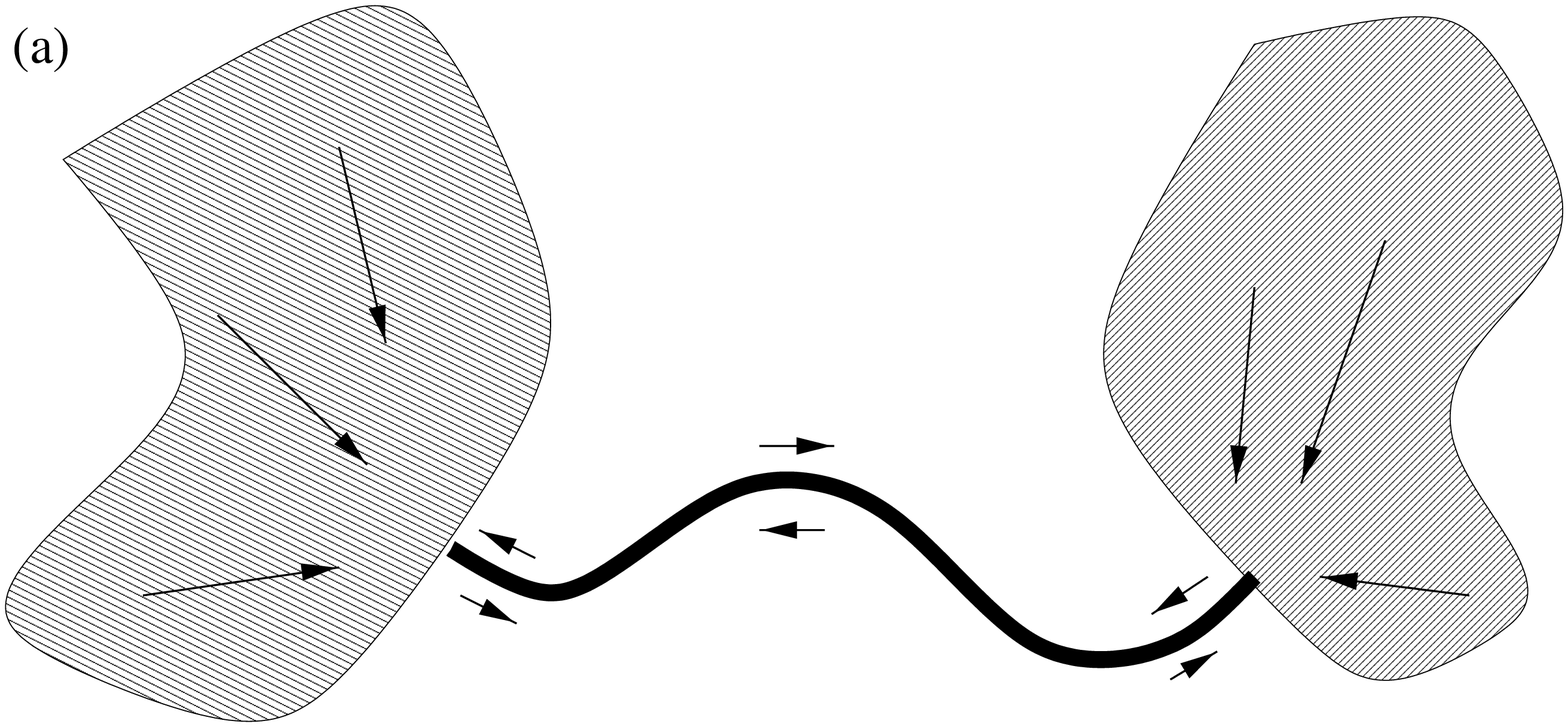} 
\vspace*{1cm}
\epsfxsize=150mm
\epsffile{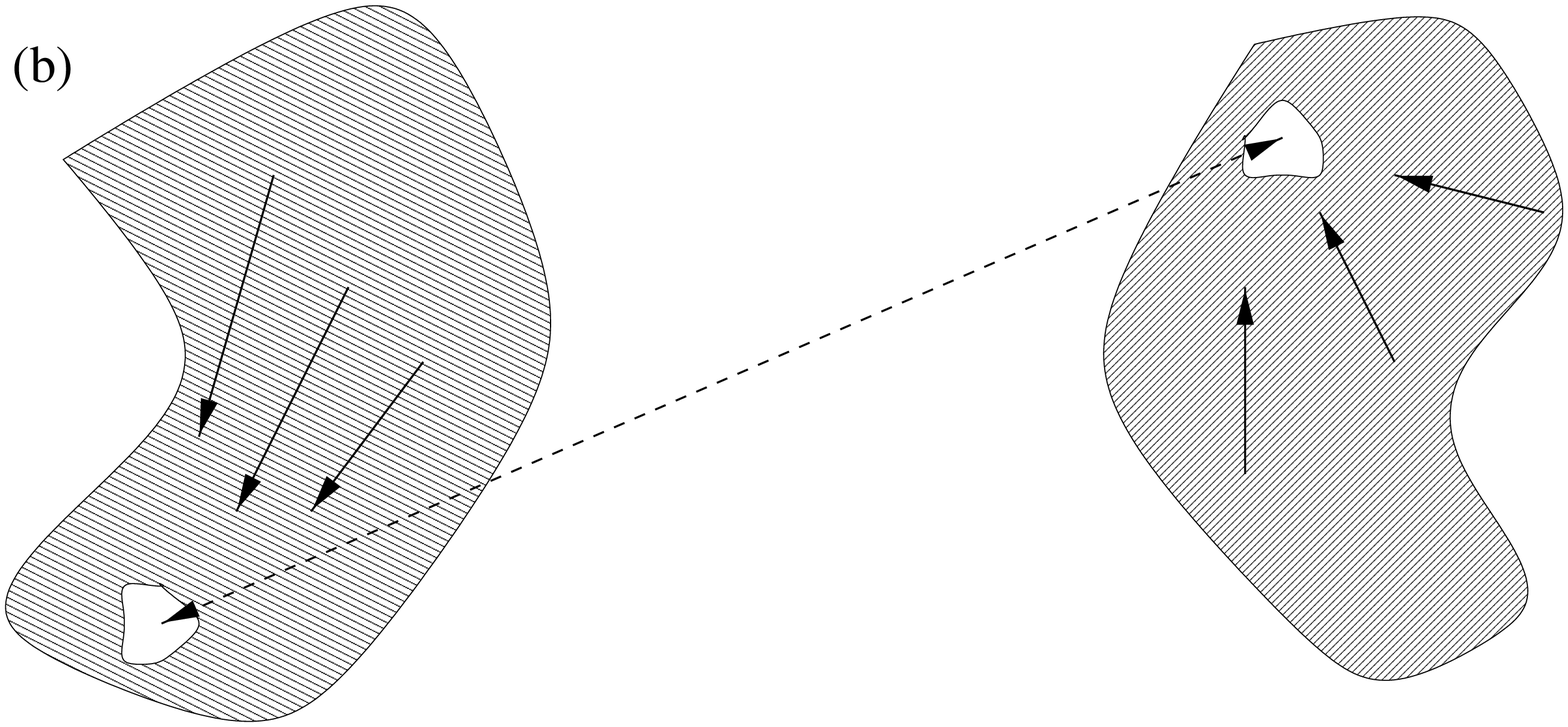} 

\vspace*{1cm}

\caption{
\label{fig:schematic}
Schematic representations of the different ways in which multicanonical
sampling methods can be used to achieve inter-phase crossing. In the
conventional approach (a) the sampling algorithm is biased so as
to enhance the probability of the {\em mixed} phase states lying along a path
(the heavy dark line) linking the two regions of configuration space. In the
lattice-switch method (b) the bias  is constructed so as to enhance the
probability of the subsets of states (the white islands), {\em within} the
single-phase  regions, from which the switch operation (the large dashed arrow)
will be accepted.
}

\end{figure}

\newpage

\begin{figure}[h]
\epsfxsize=160mm
\epsffile{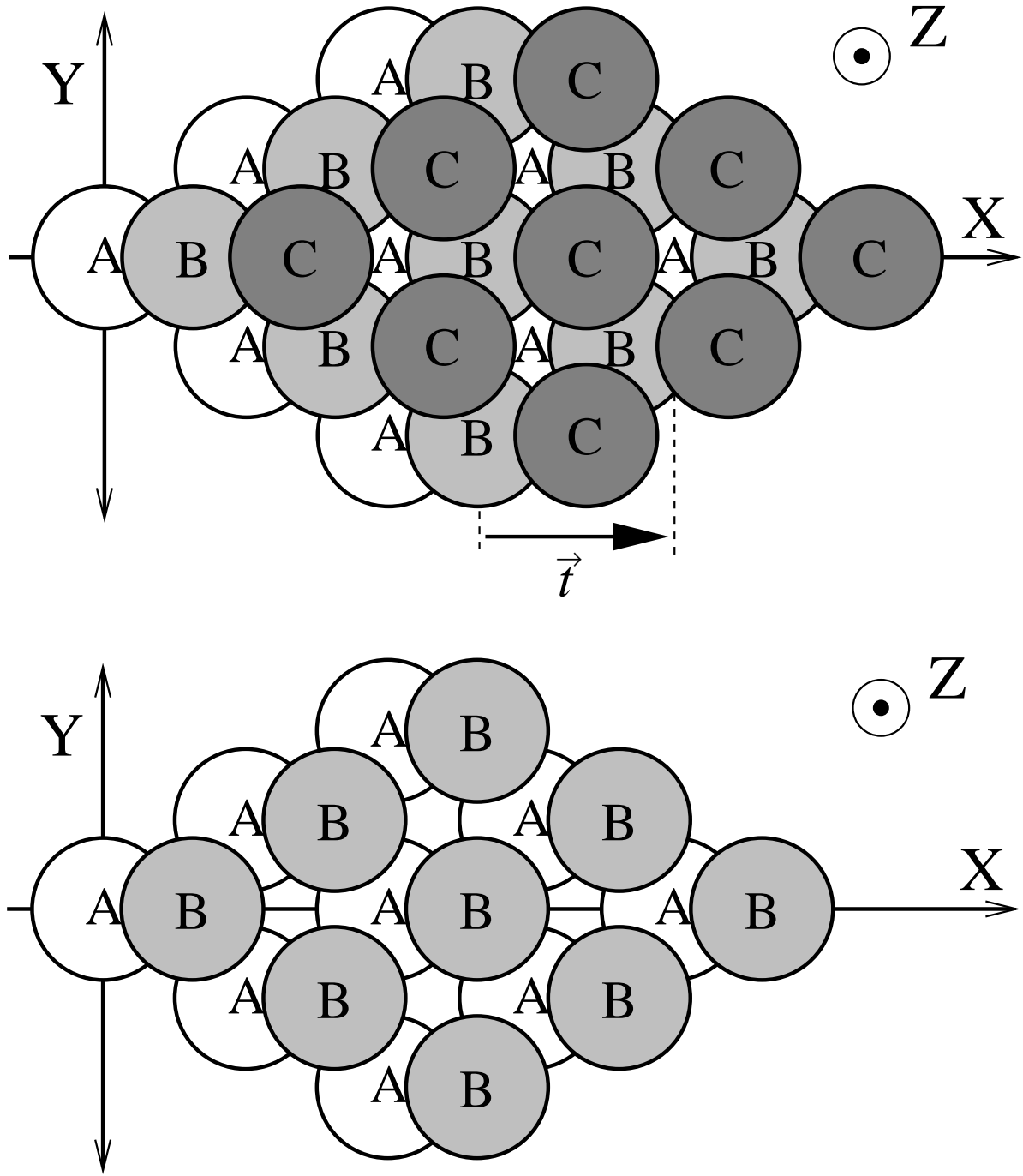}

\vspace*{1cm}

\caption{
\label{fig:stackpatt}
Schematic representations of the two close-packed structures.
The structures differ only in regard to the stacking pattern of
the close-packed ($x$-$y$) planes which are of the form 
$ABCABC\ldots$ for {\it fcc} (upper) and $ABAB\ldots$ for {\it hcp} (lower).
The vector labeled $\vec{t}$ 
is instrumental in defining the LS operation, shown in
 Fig.~(\ref{fig:latticeswitch}).}
\end{figure}
\newpage

\begin{figure}[h]
 \epsfxsize=160mm
\epsffile{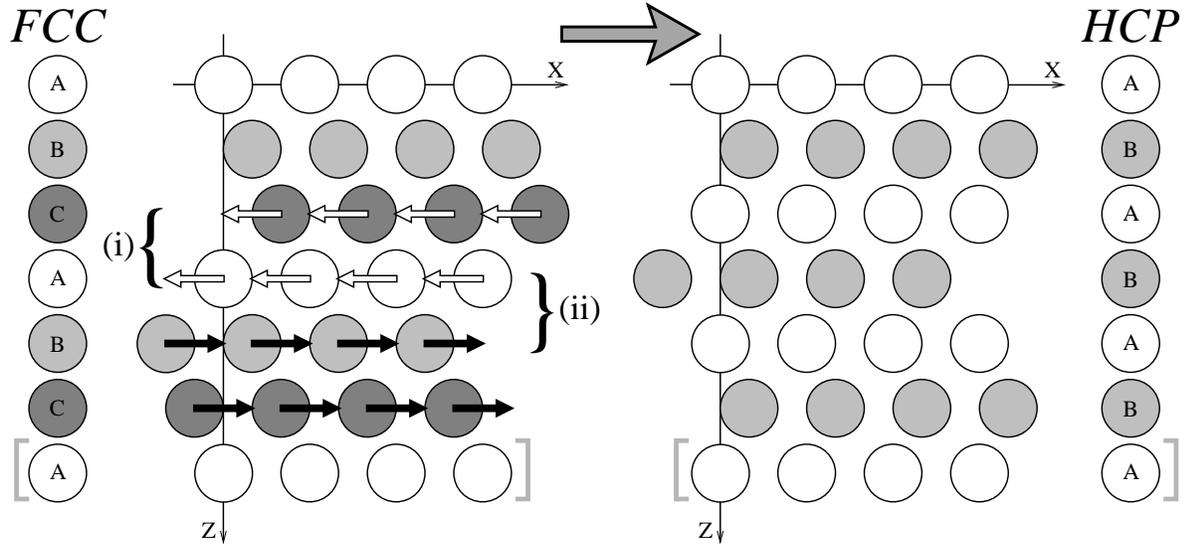}

\vspace*{1cm}

\caption{
\label{fig:latticeswitch}
The LS transformation applied to the {\it perfect-crystal} configuration.
The diagram shows 6 close-packed ($x-y$) layers. [The additional bracketed layer
at the bottom is the periodic image of the layer at the top.]
The circles show the boundaries of hard spheres located 
at the sites of the two  close-packed structures. 
In this realization of the  $fcc \rightarrow hcp$
lattice-switch, the top pair of planes are left unaltered, while the other pairs
of planes are relocated by translations, specified by the vectors 
$-\vec{t}$ (white arrows) and
$\vec{t}$ (black arrows). The vector $\vec{t}$ is identified 
in Fig.~(\ref{fig:stackpatt}).
}
\end{figure}
\newpage

\begin{figure}[h]
 \epsfxsize=160mm
\epsffile{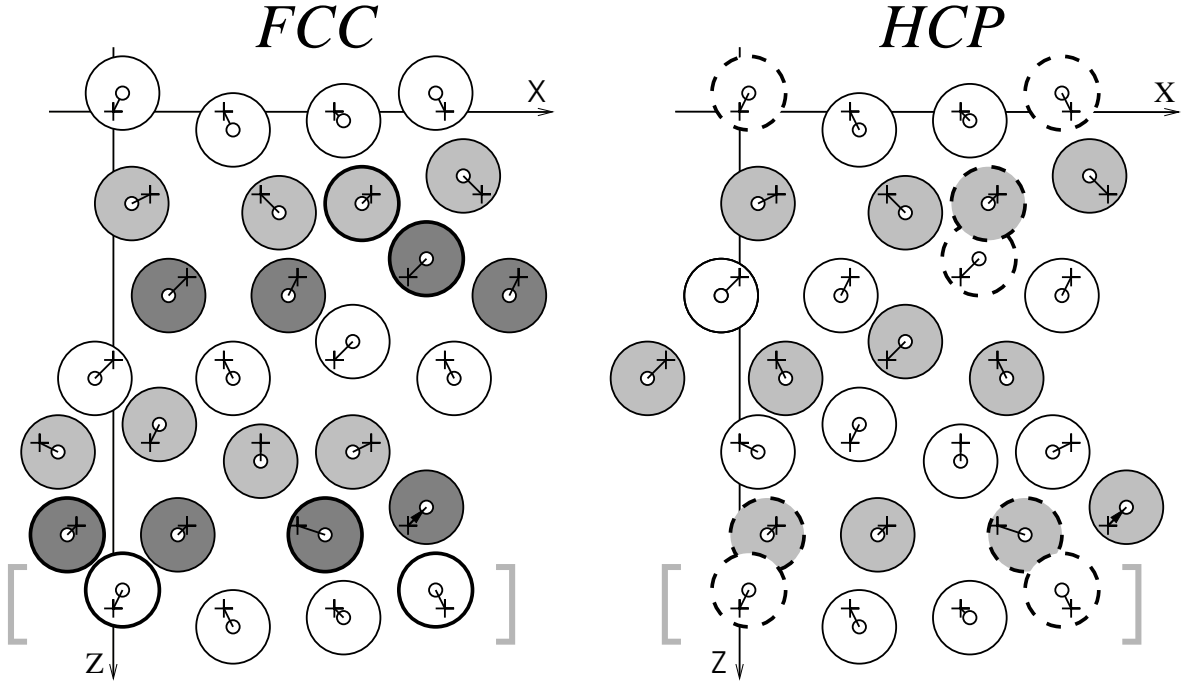}

\vspace*{1cm}

\caption{
\label{fig:orderparam}
The LS transformation applied to a `typical' configuration.
The crosses identify the `lattice sites'; the small circles locate the sphere
centers in this configuration of displacements
$\{\vec{u}\}$. This configuration is realizable (gives
no overlaps) in the $fcc$ structure; under the LS transformation it
is mapped onto an (unrealizable) $hcp$ configuration with
four overlapping pairs of hard spheres (shown with dashed boundaries).  Thus, for this configuration,
the overlap order parameter ${\mathcal M}(\{\vec{u}\}) = 4$ 
(Eq.~\ref{eq:overlapdef}).
}
\end{figure}

\newpage

\begin{figure}[h]
 \epsfxsize=160mm
\epsffile{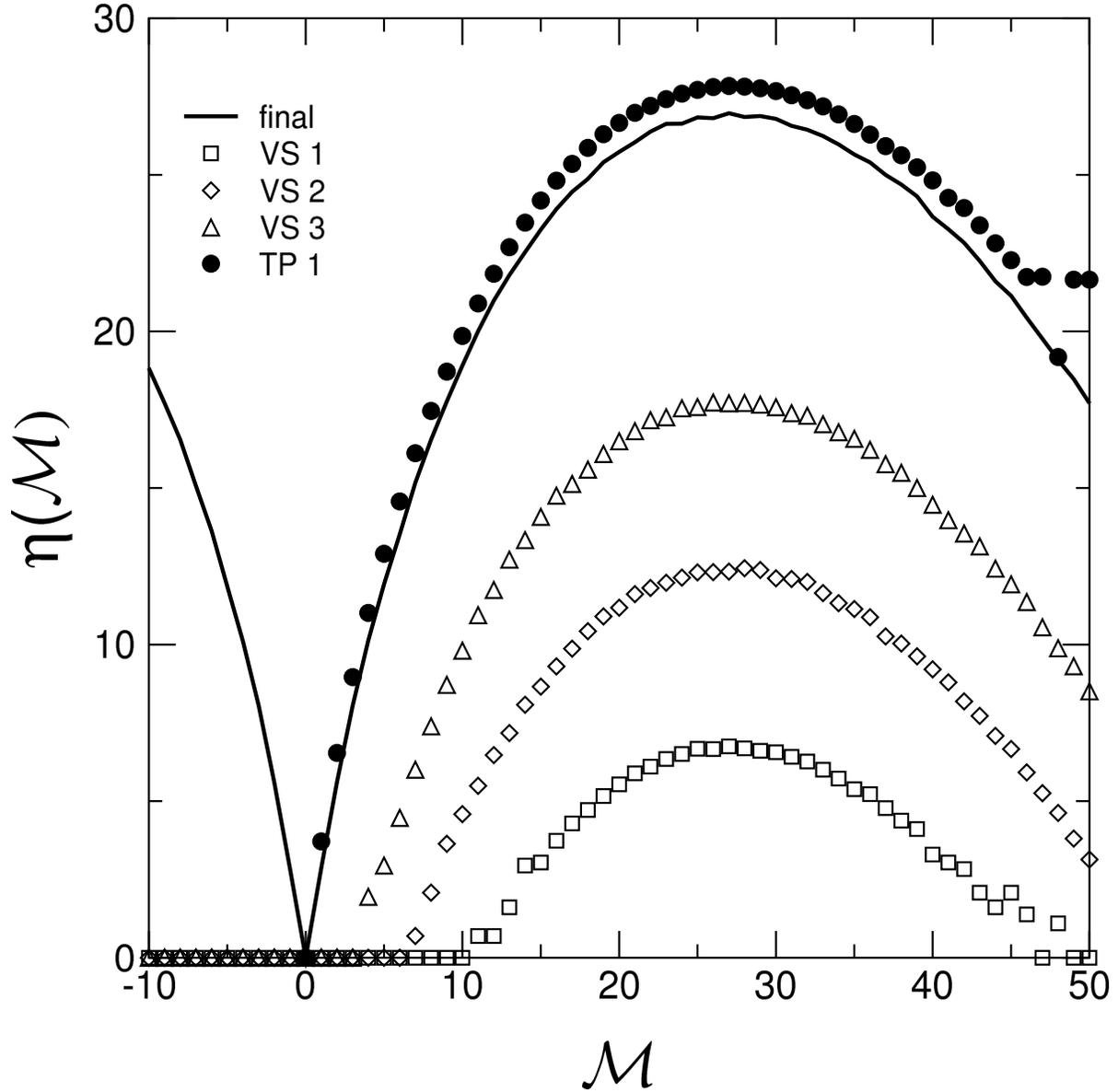}

\vspace*{1cm}

\caption{
\label{fig:weightgeneration}
Illustration of the weight-generation process, for a system of $N=216$
hard spheres. The points marked VS are the results of the first 3 iterations of
the visited-states algorithm, initiated from an {\it fcc} equilibrium state.
The points marked TP emerge from one application of the transition probability
method. The solid line shows a refined (usable) set of weights.}
\end{figure}

\newpage
\begin{figure}[h]
\epsfxsize=160mm
\epsffile{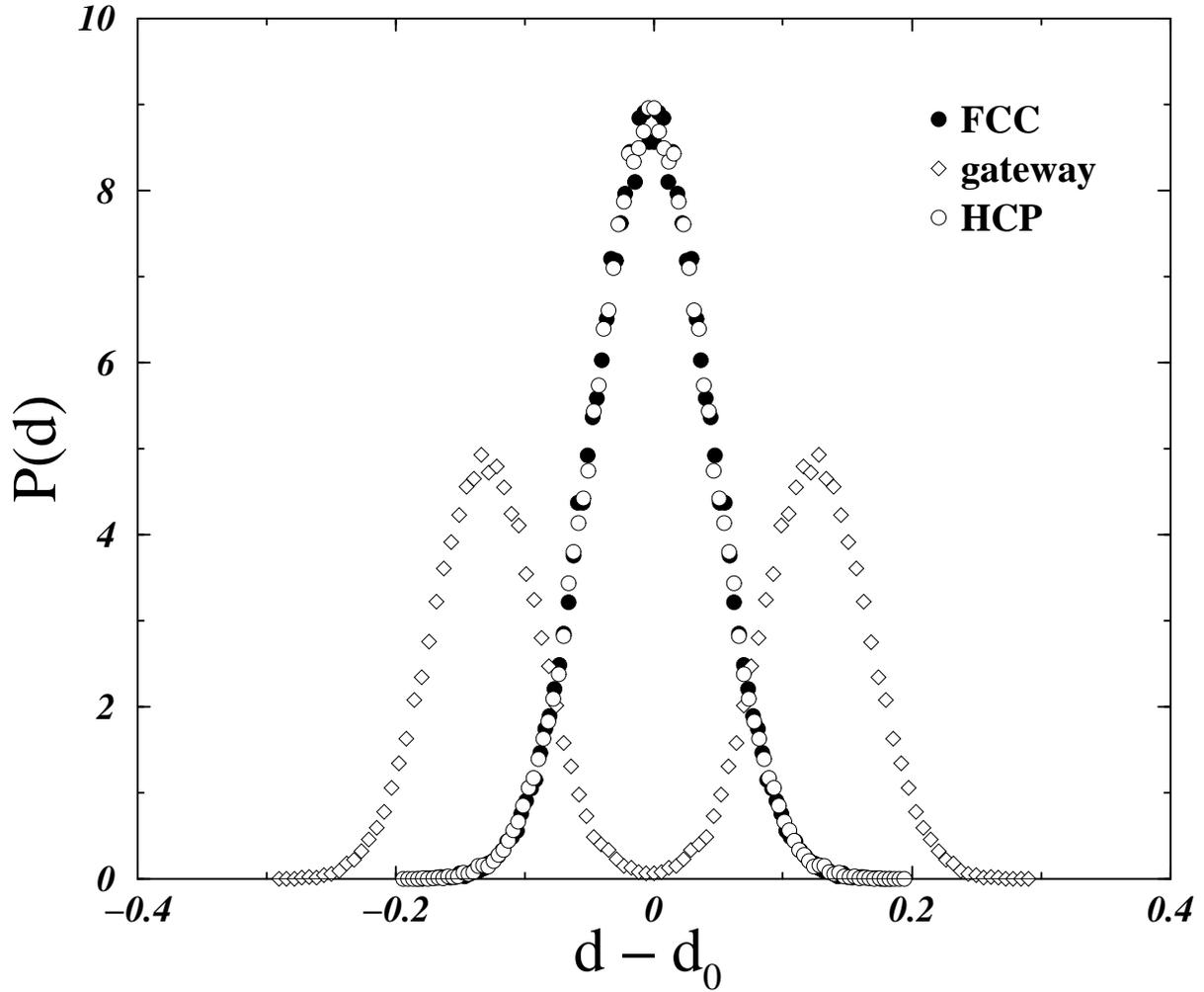}

\vspace*{1cm}

\caption{
\label{fig:sepdist}
Distribution of the separation $d$ between adjacent close-packed planes
in a system of 216 spheres at $\tilde{\rho}=0.7778$, in the equilibrium {\it hcp} and {\it fcc} macrostates,
and in the gateway (${\cal M}=0$) macrostate.
The separation is measured with respect to the equilibrium separation $d_0$ and
is expressed in units of the sphere separation $\delta$ \protect\cite{deltadef}.
}
\end{figure}
\newpage

\begin{figure}[h]
\begin{tabular}{cc}
(a) & (b)\\
\leavevmode
\epsfxsize=70mm
\epsffile{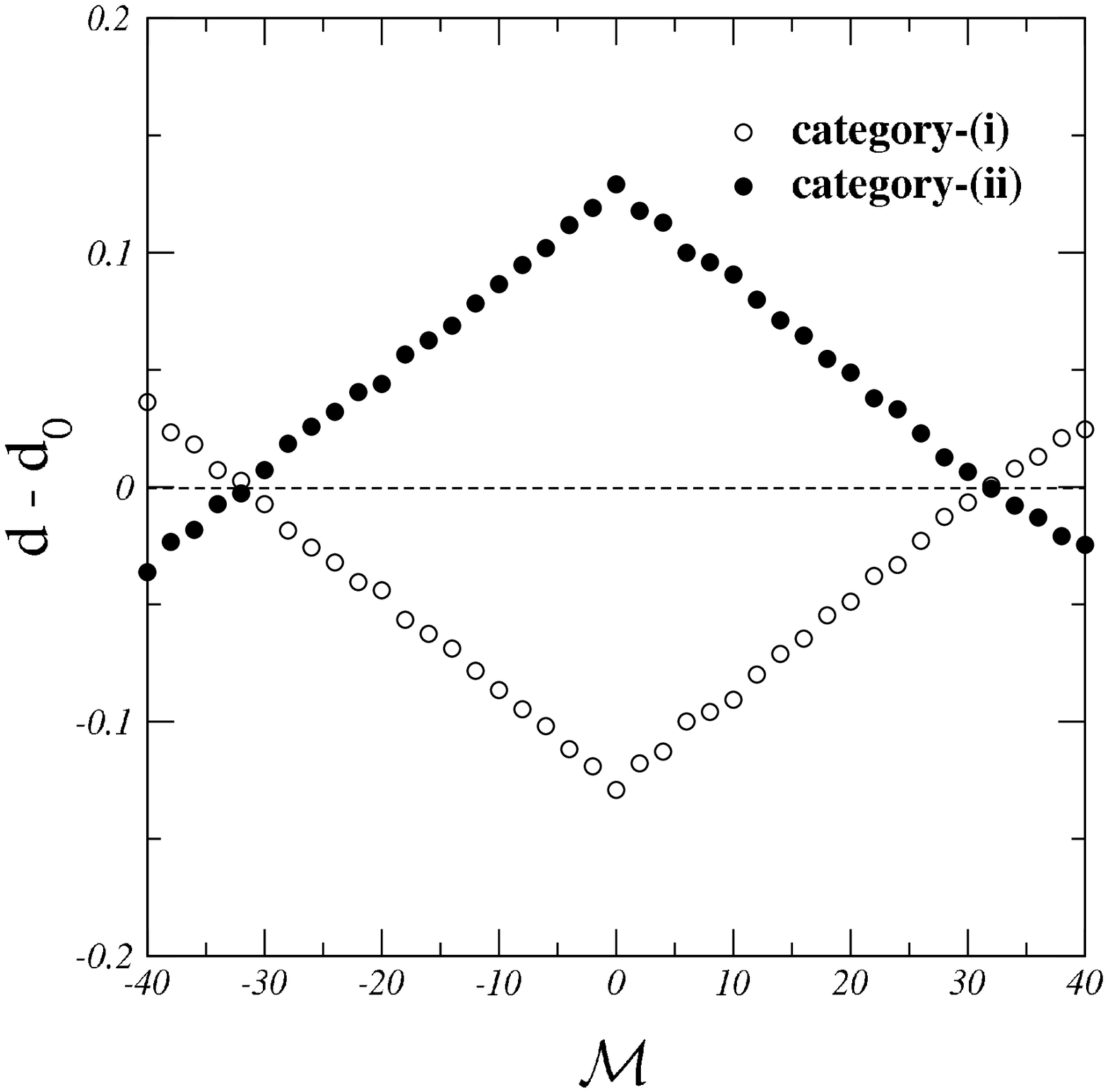}\hspace*{4mm}&
\epsfxsize=70mm
\epsffile{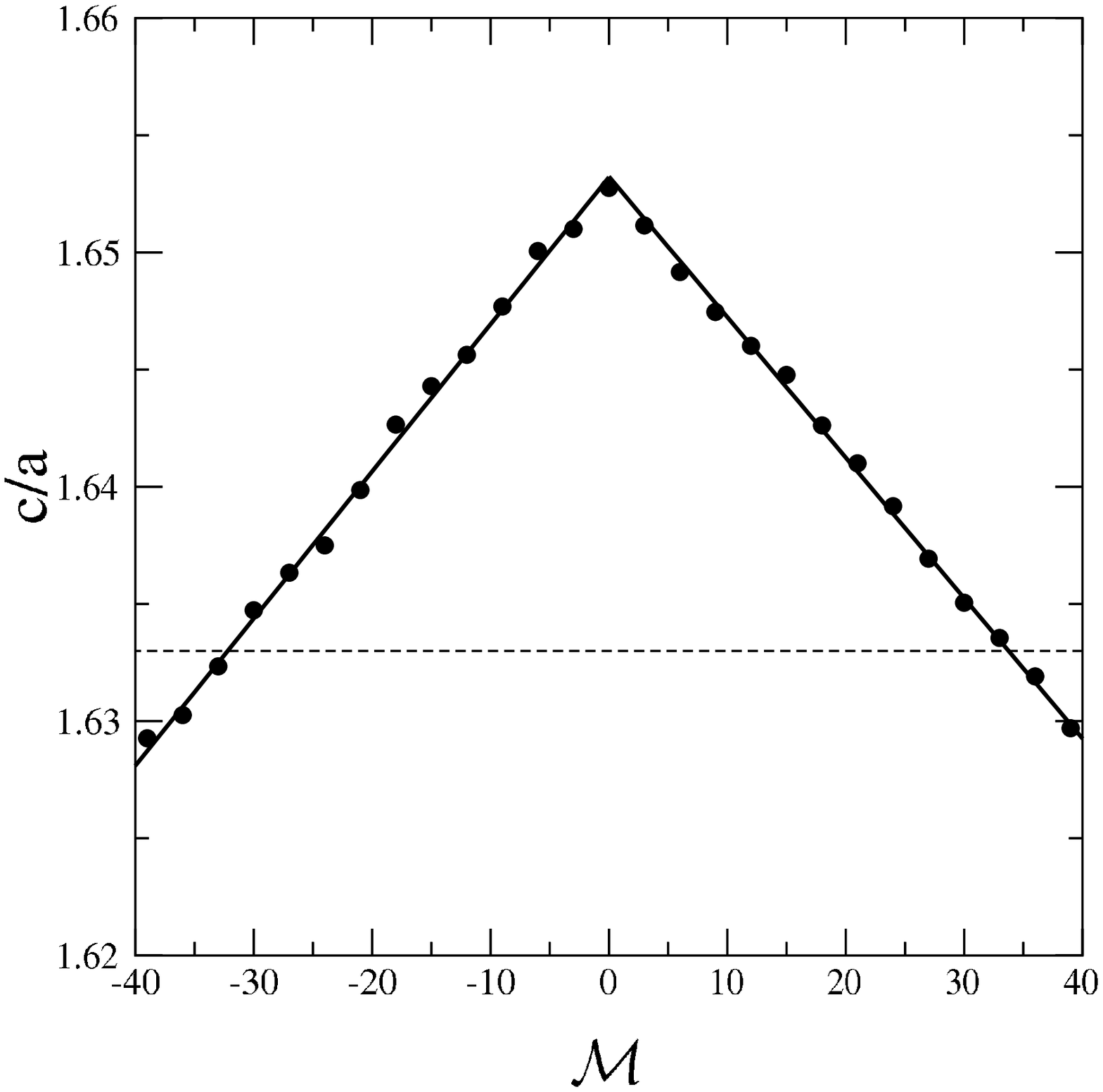}
\end{tabular}
\vspace*{1cm}

\caption{
\label{fig:gatewayfeatures}
(a) The mean value of the separation $d$ between adjacent close-packed planes in
a system of 216 spheres at $\tilde{\rho}=0.7778$, for macrostates of different
${\cal M}$ . The separation is measured with respect to the equilibrium separation
$d_0$  in units of $\delta$ \protect\cite{deltadef}.  Category-(i) planes
(see Fig.~\ref{fig:latticeswitch}) are translated together by LS;  category-(ii)
planes are translated through different amounts by LS. \\
(b) The evolution 
with ${\cal M}$ of the $c/a$-ratio
\protect\cite{ca-ratio} in a constant-{\em pressure} ensemble (with the same
parameters as (a)). The horizontal line marks the ideal-close-packed value.  
}

\end{figure}
\newpage

\begin{figure}[h]
 \epsfxsize=160mm
\epsffile{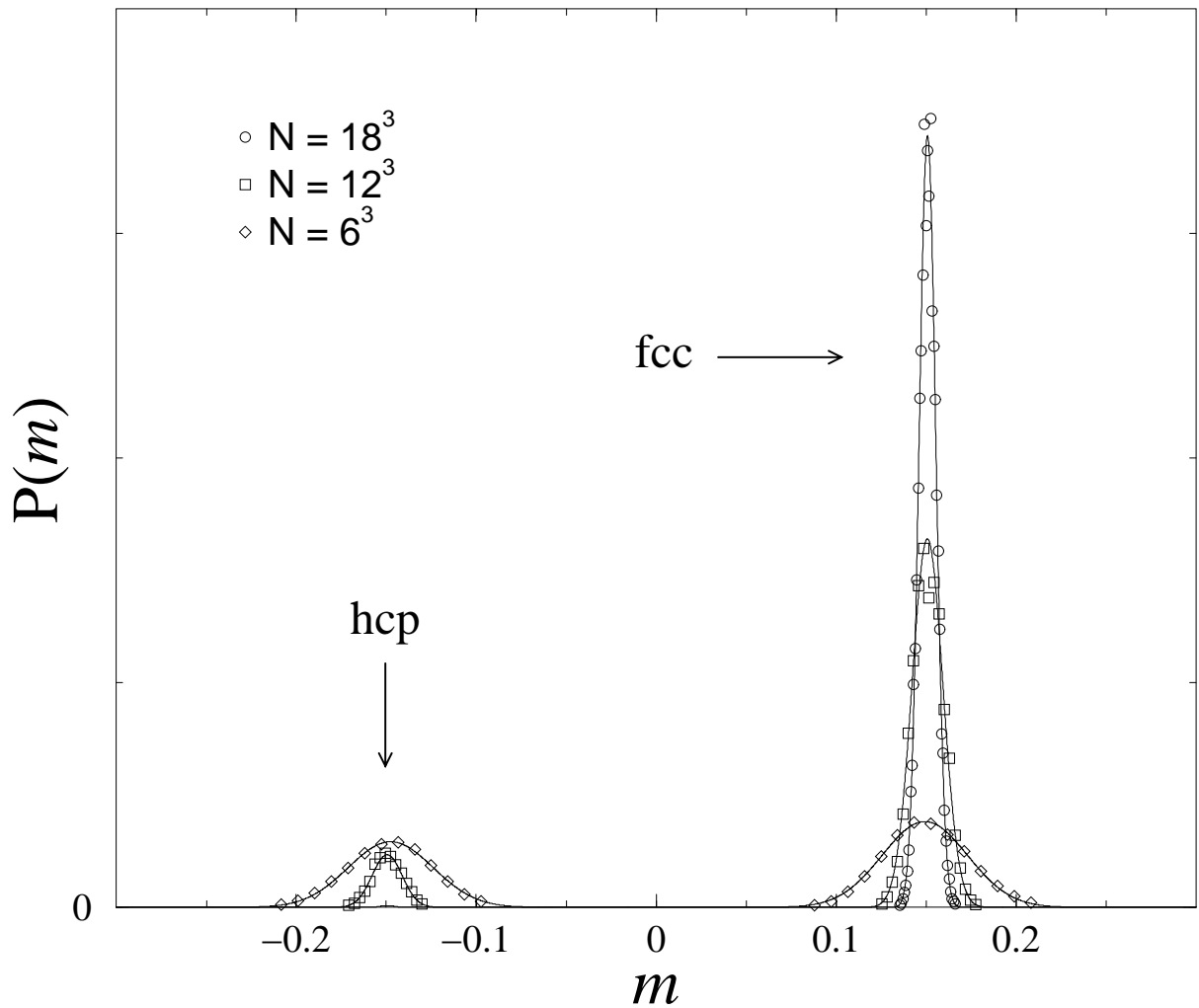}

\vspace*{1cm}

\caption{
\label{fig:pvsmvarious}
The probability distribution of the overlap order parameter per particle, $m\equiv {\cal M}/N$, 
for systems of three different $N$ values at $\tilde{\rho}=0.7778$.
The lines provide gaussian guides to the eye; the
statistical
uncertanties on the data points are smaller than the symbol size. 
The entropy-difference
is identified from the logarithm of the ratio of the
integrated weights of the two peaks. The {\it hcp} peak for the largest system
is not visible on this scale.}
\end{figure}

\newpage
\begin{figure}[h]
 \epsfxsize=160mm
\epsffile{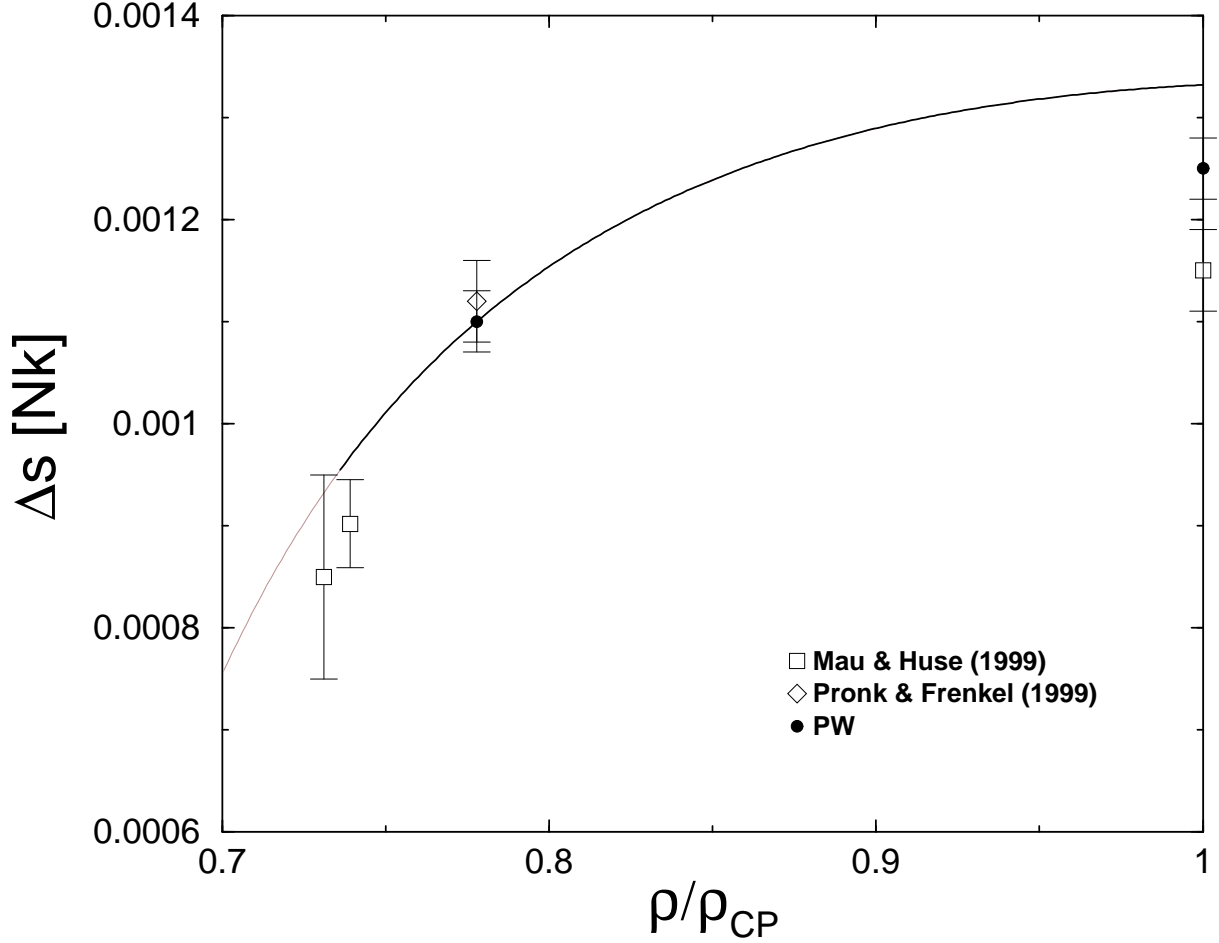}

\vspace*{1cm}

\caption{
\label{fig:speedycurve}
The difference in the entropy
densities 
of the $fcc$ and $hcp$ structures,
$\Delta s\equiv \Delta s_{\mbox{\it fcc,hcp}}$ (Eq.~\protect\ref{eq:entdiff}),
as a function of reduced density $\tilde{\rho}$.
The data points are
as given in Table \protect\ref{tab:sdiff}. The solid line is the result of an
integration of the pressures of the phases \protect\cite{speedy}. Note that
this line passes through our result at $\tilde{\rho}=0.7778$
by {\em construction}. 
}
\end{figure}

\bibliographystyle{/Disk/teTeX/texmf/tex/latex/physics/revtex/prsty}

\bibliography{refs}

\begin{references}
\bibitem{ormoldyn} Most, perhaps all, of the methods referred to here may also
be realized within the framework of molecular dynamics.
\bibitem{howtolabelpath} A  path is defined by a sequence of values of either
 some macroscopic observable  or
 some  parameter, $\lambda$, which controls a thermodynamic field
 or a model parameter.
\bibitem{notobviousapriori}
 The nature of the configurations visited along a path is not
 generally predictable {\em a priori} from the choice of order-parameter.
 One may say only that the configurations sampled at a given point 
 on the path will be those which have  measurable {\em canonical} probabilities
 {\em conditioned} on that macrostate, and which are, moreover,
 accessible on the relevant timescales.
\bibitem{interface} The interfacial free energy emerges as a by-product---and
 in some cases \cite{bergint} may actually be the principal focus of interest.
\bibitem{notbravaislattice} The hcp `lattice' is not a `Bravais lattice'; see eg
reference \cite{ashmerm} p79.
\bibitem{classartifact} The divergence is an artifact of the
   classical character of the model.
\bibitem{hardtosee} The `visualization' exercise has its limitations. The
  dodecahedra (and the lattice spacing) should be thought of as `infinitely 
large'
  compared to the mean separation of adjacent faces.
\bibitem{units} We subsume a factor of $k_BT$ into the `pressure' $P^{*}$ .
\bibitem{signconvention} The sign convention here has no deep significance.
  Defining ${\cal M}$ so that it has different signs in the two phases simply allows us to
  make a visually clear distinction (Fig.~\ref{fig:pvsmvarious}) between the
  contributions which each phase makes to the ${\cal M}$-distribution.
\bibitem{whythisdensity} This density value was chosen to coincide with one of
 those studied in Ref~\cite{frenkelladd}.
\bibitem{sortofmauhuse} This conclusion is in qualitative accord with that of
reference \cite{mauhuse}, although we find the effects of next-neighbor
encounters to be somewhat smaller than is reported there.
\bibitem{whatissimple} By `simple' we mean a $T$ matrix that preserves locality,
merely acting to rotate or reflect individual displacements. 
Non-local transformations which mix the displacement vectors might
reduce the overlap count further.
\bibitem{morecleverstill} The gateway configurations have further distinctive
features (with respect to their equilibrium counterparts): the root-mean-square
particle displacement, in the z-direction,  is reduced; and the  in-plane
correlation length of these displacements (the undulations of the close packed
planes) is enhanced along the $\vec{t}$-direction.
\bibitem{deltadef} We define $\delta = D (\tilde{\rho} ^{-1/3} -1)$, giving
the shortest distance between two sphere surfaces in the ideally-close-packed
perfect crystal configuration.
\bibitem{butnotquite} A gaussian parameterisation of the
peaks provides a first approximation to a set of multicanonical weights---but
one which significantly underestimates the entropic barrier against the LS.
\bibitem{shearmethod} The  `shear' implementation described by Mau and Huse
  \cite{mauhuse} is (they report) substantially harder to
  implement than the switch method which we discuss here, and which they use in most of
  their work.
\bibitem{butsign} But $\Delta s$ and $\Delta g$ are of opposite sign.
\bibitem{notstackinglabel} Note our double use of $\alpha$ as
 both a stacking-type-probability  and a structure label.
\end{references}

\end{document}